\documentclass[12pt]{article}
\pdfoutput=1
\usepackage{epsf,color,colordvi}
\usepackage{graphicx}
\usepackage{amsmath}
\usepackage{amssymb}
\usepackage{epsfig}
\usepackage{cite}
\usepackage{fontenc}
\usepackage{float}
\usepackage[lofdepth,lotdepth,caption=false]{subfig}
\usepackage{color}
\usepackage{booktabs}
\usepackage{tabularx}
\usepackage{multirow}
\usepackage{longtable}
\usepackage{lscape}
\usepackage{dcolumn}
\usepackage{rotating}
\usepackage{hyperref}
\usepackage[normalem]{ulem}

\newcommand{\tc}{\theta_{23}}

\definecolor{darkgreen}{cmyk}{1,0,1,0.4}
\definecolor{brown}{cmyk}{0,0.8,1,0.2}
\definecolor{darkred}{cmyk}{0,1,1,0.2}

\def\hlpm#1{\textcolor{black}{\textrm{#1}}}

\allowdisplaybreaks[4] 
\makeatletter
\renewcommand{\fnum@table}{\textbf{\tablename~\thetable}}
\renewcommand{\fnum@figure}{\textbf{\figurename~\thefigure}}
\makeatother

\newcounter{myenumi}

\renewcommand{\themyenumi}{\roman{myenumi}}
{\end{list}}
\newlength{\myem}
\newcounter{mysubequation}[equation]

\makeatletter
\renewcommand{\section}{\@startsection{section}{1}{0em}{-\baselineskip}%
{\baselineskip}{\normalfont\large\bfseries}}
\renewcommand{\subsection}%
{\@startsection{subsection}{2}{0em}{-0.7\baselineskip}%
{0.7\baselineskip}{\normalfont\bfseries}}
\makeatother
\newcommand{\bi}{\begin{itemize}}
\newcommand{\ei}{\end{itemize}}

\def\beq{\begin{equation}}
\def\eeq{\end{equation}}
\newcommand{\bea}{\begin{eqnarray}}
\newcommand{\D}{\Delta}
\newcommand{\eea}{\end{eqnarray}}
\newcommand{\Ap}{\hat{A}}

\newcommand{\ldm}{\Delta m_{31}^2}
\newcommand{\sdm}{\Delta m_{21}^2}

\def\epsilon{\varepsilon}

\newcommand{\chisq}{\ensuremath{\chi^2}}

\newcommand{\dune}{{\sc DUNE}}

\def\<{\langle}
\def\>{\rangle}

\def\dfrac#1#2{{\displaystyle\frac{#1}{#2}}}

\def\lsim{\mathrel{\rlap{\lower4pt\hbox{\hskip1pt$\sim$}}
    \raise1pt\hbox{$<$}}}         
\def\gsim{\mathrel{\rlap{\lower4pt\hbox{\hskip1pt$\sim$}}
    \raise1pt\hbox{$>$}}}         

\newcommand{\dacp}[1]{\ensuremath{\delta [\Delta P^{CP/T}_{\alpha\beta}]}}

\newcommand{\pbarab}[1]{\ensuremath{{ P}_{\bar{\alpha} \bar{\beta}} }}
\newcommand{\pbarba}[1]{\ensuremath{{ P}_{\bar{\beta} \bar{\alpha}} }}
\newcommand{\acpab}[1]{\ensuremath{A^{CP}_{\alpha \beta}}}
\newcommand{\acpaa}[1]{\ensuremath{A^{CP}_{\alpha \alpha}}}
\newcommand{\ataa}[1]{\ensuremath{A^{T}_{\alpha \alpha}}}
\newcommand{\acpba}[1]{\ensuremath{A^{CP}_{\beta \alpha}}}
\newcommand{\atab}[1]{\ensuremath{{A}^{T}_{\alpha \beta}}}
\newcommand{\atba}[1]{\ensuremath{{A}^{T}_{\beta \alpha}}}

\newcommand{\acptab}[1]{\ensuremath{A^{CPT}_{\alpha \beta}}}
\newcommand{\dpcpab}[1]{\ensuremath{\Delta {\cal P}^{CP}_{\alpha \beta} }}
\newcommand{\dptab}[1]{\ensuremath{\Delta {\cal P}^{T}_{\alpha \beta}}}
\newcommand{\dpcptab}[1]{\ensuremath{\Delta {\cal P}^{CPT}_{\alpha \beta} }}

\begin{document}
\begin{titlepage}

\renewcommand{\thefootnote}{\alph{footnote}}

\vspace*{-3.cm}
\begin{flushright}

\end{flushright}


\renewcommand{\thefootnote}{\fnsymbol{footnote}}
\setcounter{footnote}{-1}

{\begin{center}
{\Large\bf 
Physics prospects with the second oscillation 
 maximum at Deep Underground Neutrino Experiment 
\\
}
\end{center}}

\renewcommand{\thefootnote}{\alph{footnote}}

\vspace*{.8cm}
\vspace*{.3cm}
{\begin{center} 
            {\sf    
                 Jogesh Rout${^{\S \,}}$\,\footnote[1]
                {\makebox[1.cm]{Email:}  jogesh.rout1@gmail.com},
                 Sheeba Shafaq$^\S$\,\footnote[2]
                 {\makebox[1.cm]{Email:} sheebakhawaja7@gmail.com},
                 Mary Bishai${^{\P \,}}$\,\footnote[3]
                {\makebox[1.cm]{Email:}  mbishai@bnl.gov}, and
                Poonam Mehta$^{\S}$\,\footnote[4]
                {\makebox[1.cm]{Email:} pm@jnu.ac.in}
              
}

\end{center}
}
\vspace*{0cm}
{\it 
\begin{center}
$^\S$\, School of Physical Sciences,      Jawaharlal Nehru University, 
      New Delhi 110067, India  \\
$^\P$\, Brookhaven National Laboratory,    PO Box 5000
Upton, NY 11973-5000

\end{center}}


\begin{center}
{ \today}
\end{center}

{\Large 
\bf
 \begin{center} Abstract  
\end{center} 
 }
Current long-baseline neutrino-oscillation experiments such as
    NO$\nu$A and T2K are mainly sensitive to physics in the
    neighbourhood of the {\sl{first oscillation maximum}} of the
    $\nu_\mu \to \nu_e$ oscillation probability. The future Deep
    Underground Neutrino Experiment (DUNE) utilizes a wide-band beam
    tune optimized for CP violation sensitivity that fully covers the
    region of the first maxima and part of the second.  In
    the present study, we elucidate the role of {\sl{second oscillation
        maximum}} in addressing issues pertaining to unknowns in the standard three flavour paradigm. 
        We consider  a new DUNE beam tune optimized for
    coverage of the region of the second oscillation maxima which could
    be realized using proposed accelerator upgrades that provide multi-MW
    of power at proton energies of 8 GeV. 
     We find that addition of the multi-MW 8 GeV beam to DUNE wide-band running leads to modest improvement  in
    sensitivity to CP violation, mass hierarchy, the octant of $\theta_{23}$ as well as the resolution of $\delta$ and the Jarlskog invariant. 
    Significant improvements to the DUNE neutrino energy resolution yield a much larger improvement in performance.
    We conclude that the standard DUNE wide-band beam when coupled with excellent detector
    resolution capabilities is sufficient to resolve $\delta$ to
    better than $\sim 12^\circ$ for all values of $\delta$ in a decade
    of running.  For second maxima    (8 GeV 3MW) beam  running concurrently with the standard wide-band   (80 GeV 2.2 MW) beam
     for 5 of the 10 years, it is found that $\delta$ can be further resolved better than $\sim 10^\circ$ for all
    values of $\delta$.

\vspace*{.5cm}

\end{titlepage}

\newpage

\renewcommand{\thefootnote}{\arabic{footnote}}
\setcounter{footnote}{0}

\section{Introduction}
\label{sec:intro}

Pontecorvo's original insight that neutrinos oscillate among one another~\cite{pontecorvo} has been confirmed by a variety of neutrino oscillation experiments involving wide range of energies and baselines. The idea of neutrino oscillations among the three light active neutrino flavours has been rewarded with a Nobel prize in 2015~\cite{nobel2015}.
The parameters entering the neutrino oscillation framework have been measured to a fairly good precision (see the recent global fit analyses~\cite{deSalas:2020pgw,Esteban:2020cvm}). The best-fit values and $3\sigma$ range of neutrino mass and mixings deciphered from oscillation data are given in Table~\ref{tab:parameters}. Yet, there are some open questions in the standard mass-induced oscillation framework. 
These include the question of neutrino mass hierarchy (sign of $\Delta m^{2}_{31}$), the value of the CP violating phase ($\delta$) and determining the correct octant of $\theta_{23}$. Furthermore, it is desirable to improve the  precision measurements of the parameters  entering the oscillation framework.  

Determination of neutrino mass hierarchy (MH) would allow us to get closer towards determining the underlying structure of the neutrino mass matrix by being able to discriminate between theoretical models giving rise to neutrino masses~\cite{Albright:2006cw}. Along with the CP violating phase $\delta$, it impacts the effectiveness of leptogenesis scenario which can explain the matter-antimatter asymmetry of the Universe~\cite{Fukugita:1986hr}.

The next generation neutrino oscillation experiments would allow us to precisely determine the known parameters and determine the remaining unknowns in the neutrino oscillation formalism. The long-baseline neutrino experiments are designed such that the desirable physics outcome is achieved. Typically, the optimal combination is for a value of baseline ($L$) and  energy ($E$) for which $P_{\mu e}$ has its first peak. This is referred to as the {\sl{first oscillation maximum}}. Typically, for shorter baselines, the higher oscillation maxima are unaccessible as the energies at which these  occur are low and difficult to produce experimentally . For longer baselines, it may be possible to access the information from the {\sl{second (and higher) oscillation maxima}}. At the second oscillation maximum, one expects higher sensitivity to $\delta$ as the size of the $\delta$-dependent interference term  is a factor of $\sim 3$  larger than that at the first oscillation maximum.  

A promising future experiment is the Deep Underground Neutrino Experiment (DUNE). A high purity muon neutrino beam will be produced at Fermilab and will travel 1300 km to a liquid Argon (LAr) far detector placed at an on-axis location at Sanford Underground Research Facility (SURF). The primary aim of DUNE is to address the question of CP violation and identify the neutrino mass hierarchy~\cite{2013arXiv1307.7335L,Acciarri:2015uup,Acciarri:2016crz}.
 A wide-band neutrino beam originating from the Fermilab proton complex is considered for DUNE. 
A systematic evaluation of optimal baseline for discovery of CP violation, determination of  the mass hierarchy and resolution of  the $\theta_{23}$ octant in a long baseline oscillation experiment was carried out by Bass et al.~\cite{Bass:2013vcg} and it was concluded that for achieving unambiguous measurement of these parameters, one needed a baseline atleast of the order of $1000$ km.  It was further shown from the asymmetry plot that CP measurement was better achieved in the vicinity of second  oscillation maximum irrespective of the mass hierarchy and results for sensitivities to standard three flavour oscillation parameters were presented.   
 The authors had considered two detector types - Water Cherenkov (WC) and LAr and performed the study for the 
erstwhile Long Baseline Neutrino Experiment (LBNE).

 \begin{table}[h]
\centering
\scalebox{0.9}{
\begin{tabular}{| c | c | c | c |}
\hline
&&&\\
Parameter & Best-fit value & 3$\sigma$ range & $3\sigma$ uncertainty  \\
&&& [\%]\\
\hline
&&&\\
%
$\theta_{12}$ [$^\circ$]             & 34.3                    &  31.4 - 37.4   &  8.72 \\
$\theta_{13}$ (NH) [$^\circ$]    & 8.58              &  8.16  -  8.94   & 4.56 \\
$\theta_{13}$ (IH) [$^\circ$]    & 8.63              &  8.21  -  8.99   & 4.53 \\
$\theta_{23}$ (NH) [$^\circ$]        & 48.8                     &  41.63  - 51.32    & 10.42  \\
$\theta_{23}$ (IH) [$^\circ$]        & 48.8                     &  41.88  - 51.30    &   10.11 \\
$\sdm$ [$\text{eV}^2$]  & $7.5 \times 10^{-5}$  &  [6.94 - 8.14]$\times 10^{-5}$  &  8.0 \\
$\ldm$ (NH) [$\text{eV}^2$] & $+2.56 \times 10^{-3}$   &  [2.46 - 2.65] $\times 10^{-3}$ & 3.72  \\
$\ldm$ (IH) [$\text{eV}^2$] & $-2.46 \times 10^{-3}$  & -[2.37 - 2.55]$\times 10^{-3}$  &3.7 \\
$\delta$ (NH) [Rad]   & $-0.8\pi$   & $[-\pi, 0]  \cup [0.8\pi, \pi]$ &  $-$ \\
$\delta$ (IH) [Rad]   & $-0.46\pi$   & $[-0.86\pi, -0.1\pi]$   & $-$  \\
&&&\\
\hline
\end{tabular}}
\caption{\label{tab:parameters}
 Standard oscillation parameters and their uncertainties used in our study. The values were taken from the global fit analysis in \cite{deSalas:2020pgw}. 
 If the $3\sigma$ upper and lower limit of a parameter is $x_{u}$ and $x_{l}$ respectively, the $3\sigma$ uncertainty is $(x_{u}-x_{l})/(x_{u}+x_{l})\%$.}
\end{table}

The idea of utilizing the second oscillation maximum in neutrino experiments is not new.  The prospect of using a high intensity low energy neutrino beam using Project X was studied in~\cite{Bishai:2013yqo} and it was demonstrated that the  simultaneous operation of 8 GeV and 60 GeV beams in conjunction with a Water Cherenkov detector allows for senstivity to $\nu_\mu\to\nu_e$ oscillation at the second oscillation maximum. The focus of the study  was to attain high precision measurement of $\theta_{13}$ and $\delta_{}$. 
With the goal of enhancing the mass hierarchy sensitivity, the authors of~\cite{Qian:2013nhp} introduced a second detector at an off-axis location (same beam was used) and obtained marginal improvement for certain values of $\delta_{}$ in the worse half plane of $\delta_{} $ values. 
In~\cite{Coloma:2014kca}, the idea of utilizing different oscillation maxima was invoked in discussing the role and interplay between appearance and disappearance channels at long baseline experiments for precision measurement of $\delta$ and $\theta_{23}$. In~\cite{Huber:2010dx}, for the considered experimental setup (Fermilab to DUSEL LBNE), it was concluded that the second maximum plays only a marginal role due to the experimental difficulties to obtain a statistically significant and sufficiently background-free event sample at low energies.  {{The impact of improved neutrino energy reconstruction capabilities at DUNE has been studied in~\cite{DeRomeri:2016qwo}. }}
The  European Spallation Source (ESS) neutrino Super Beam (ESS$\nu$SB) facility in Europe    utilizes the second oscillation peak to maximize the discovery potential to leptonic CP violation with the detector   placed at $L=540$ km. The $L/E$ is such that the second oscillation peak can naturally be exploited in order to measure oscillation parameters~\cite{Baussan:2013zcy,Wildner:2015yaa,Blennow:2019bvl,Agarwalla_2014,Chakraborty_2019,Ghosh:2019sfi,Ghosh:2020vly}. Among other experiments, the MOMENT~\cite{Cao_2014} proposal ($L = 150$ km) has similar $L/E$ as the ESS$\nu$SB. The prospect of precision measurement of $\delta$ at MOMENT has been studied in~\cite{Tang:2019wsv}.
The T2HKK~\cite{Hagiwara_2008} proposal, in which the first and the second oscillation maxima are measured with two detectors located at different sites will have the same $L/E$ range. 
More recently, invisible neutrino decay at ESS$\nu$SB has been explored in~\cite{Choubey:2020dhw} and a comparative analysis of T2HK, T2HKK and ESS$\nu$SB in the context of neutrino decay has been carried out in~\cite{Chakraborty:2020cfu}.

However, it should be noted that, a comprehensive and detailed assessment of the role of different oscillation maxima for long baseline experiment such as DUNE has not been addressed in the earlier work and this is the main motivation of the present study. In the present article, we begin with a probability level analysis highlighting the role of second oscillation maxima and its impact on the current unknowns in neutrino oscillation physics. We  also address the issue of extraction of intrinsic CP violation using the two beams and show that the second oscillation maximum helps in resolving the ambiguity in CP phase measurement. We utilize optimal beam tunes to explore the precise role of first and second oscillation maxima. The standard beam tune used in almost all the studies connected with DUNE sensitivities is derived from 80 GeV proton beam energy and this is well-suited to study oscillations in the vicinity of the first oscillation maximum. The wide-band default DUNE beam also covers a portion of the region of the second oscillation maxima. However, one needs a different source to fully utilize the second oscillation maximum - here we employ an 8 GeV beam as was proposed in previous Project X studies~\cite{Bishai:2013yqo} to harness the signal at second  oscillation maximum. We note that the Project X 8 GeV multi-MW beam could be realized by the proposed Fermilab PIP-III accelerator upgrade option utilizing a 6-8 GeV pulsed Super-conducting RF (SRF) Linac~\cite{pip,osti_1631119}. The PIP-III pulsed SRF linac could generate 4MW of power at 8 GeV of which only a few 100 kW would be needed to increase the power from the Main Injector accelerator to $\sim$ 2MW @ 80 GeV. Therefore it may be possible to utilize a Main Injector beam of $\sim$ 2MW @ 80 GeV simultaneously with a $\sim$ 3MW @ 8 GeV beam derived from the SRF linac to generate neutrino beams for DUNE.

The   present article is structured as follows.
In Sec.~\ref{framework}, we describe the basic  framework used in the present work. This includes a review of electron neutrino appearance probability in vacuum and in matter, oscillation maxima of $P_{\mu e}$, CP asymmetry, mass hierarchy (MH) asymmetry and a description of an observable for separation of intrinsic versus extrinsic contribution to the CP phase. In Sec.~\ref{expt}, we describe the experimental inputs such as beam tunes used and detector details. Sec.~\ref{event} is devoted to a discussion at the level of  event rates (with a discussion on backgrounds) at DUNE using different beam tunes. In Sec.~\ref{results}, we present our main results for sensitivities to CP, mass hierarchy, octant of $\theta_{23}$, resolution of $\delta$ and $1\sigma$ contours from a two-dimensional fit to $\tc$ and $\delta$  at the level of $\chi^2$.  We summarize our outcome in Sec.~\ref{conclude}.

\section{Framework}
\label{framework}
Neutrino oscillations have their origin in the non-zero neutrino masses and mixing among the neutrino flavors. 
The standard paradigm of neutrino oscillations involves three flavours of neutrinos which are superpositions of the mass states carrying well-defined masses.  The mixing matrix (${\cal U}$)  in the Pontecorvo-Maki-Nakagawa-Sakata (PMNS) parametrization~\cite{Beringer:1900zz}  is given by
\bea
{\mathcal U}^{} \equiv 
{\mathcal U}_{PMNS} &=& \left(
\begin{array}{ccc}
1   & 0 & 0 \\  0 & c_{23}  & s_{23}   \\ 
 0 & -s_{23} & c_{23} \\
\end{array} 
\right)   
  \left(
\begin{array}{ccc}
c_{13}  &  0 &  s_{13} e^{- i \delta}\\ 0 & 1   &  0 \\ 
-s_{13} e^{i \delta} & 0 & c_{13} \\
\end{array} 
\right)  \left(
\begin{array}{ccc}
c_{12}  & s_{12} & 0 \\ 
-s_{12} & c_{12} &  0 \\ 0 &  0 & 1  \\ 
\end{array} 
\right)  \ ,
\label{u}
 \eea 
where $s_{ij}=\sin {\theta_{ij}}, c_{ij}=\cos \theta_{ij}$ and $\delta$ is the Dirac-type CP phase. 
  Additionally, if neutrinos are Majorana, there can be two additional Majorana-type phases in the three flavour case. However, those Majorana phases play no role in neutrino oscillations~\cite{giunti}.

  In a typical long baseline experiment such as DUNE, at the source, a high purity beam of muon neutrinos ($>  90\%$ pure)
  is produced via pion and kaon decays. In principle, all the oscillation channels $\nu_\mu \to \nu_e$ ($\nu_e$ appearance), $\nu_\mu \to \nu_\mu$ ($\nu_\mu$ disappearance) and $\nu_\mu \to \nu_\tau$ ($\nu_\tau$ appearance) should be 
  accessible at DUNE. However, with the standard beam tune  of the LBNF beamline,  the $\nu_\mu \to \nu_\tau$ appearance channel does not lead to sizable event samples~\cite{PhysRevD.100.016004} and it is not possible to probe this channel. In recent studies, it has been shown that higher energy beam tunes may prove useful to probe this channel~\cite{deGouvea:2019ozk,Ghoshal:2019pab,Rout:2020cxi}. However, here, we are concerned with the standard beam tune at DUNE and the flux relevant for the $2^{nd}$ maxima, so the channels considered are : $\nu_e$ appearance channel and $\nu_\mu$ disappearance channel. In order to have a clear understanding of the role of second oscillation maximum vis-a-vis the first oscillation maximum in probing the current unknowns, 
   we briefly describe the main features of the oscillation probability for the $\nu_\mu \to \nu_e$ channel.

\subsection{Brief review of $P_{\mu e}$ in vacuum and  in matter}
\label{p}
 For propagation of neutrinos through vacuum, $P^v_{\mu e}$  can be expressed as (see~\cite{Marciano:2006uc}). 
\begin{eqnarray}
P^v_{\mu e}  &\simeq& P^v_0 + P^v_{\sin \delta} + P^v_{\cos \delta} + P^v_3~,
\label{probvacuumterms}
\end{eqnarray}
where 
\begin{subequations}
\label{PROBVACUUM}
\begin{eqnarray}
P^v_0 &=& \sin^2 \theta_{23} \sin^2 2\theta_{13} \sin^2 {\Delta_{31}} ~,\label{PROBVACUUMa} \\
P^v_{\sin\delta} &=&  \sin\delta \cos\theta_{13} \sin 2\theta_{12} \sin 2\theta_{13} \sin 2\theta_{23} \sin^2{\Delta_{31}\sin(2\alpha\Delta_{31}}) ~, \label{PROBVACUUMb}\\
P^v_{\cos\delta} &=&  \cos\delta \cos\theta_{13} \sin 2\theta_{12} \sin 2\theta_{13} \sin 2\theta_{23}
 \sin{\Delta_{31} }\sin(2\alpha\Delta_{31}) \cos\Delta_{31}   ~, \label{PROBVACUUMc}\\
P^v_3 &=& \cos^2 \theta_{23} \sin^2 2\theta_{12} \sin^2 (\alpha\Delta_{31})  ~.\label{PROBVACUUMlast}\end{eqnarray}
\end{subequations}
Here $\Delta_{31}=\Delta m_{31}^2L/4E $ and $\alpha=\Delta m_{21}^2/\Delta m_{31}^2$. $E$ is the neutrino energy in GeV and $L$ is the baseline 
in km. For antineutrinos, $\delta \to -\delta$. 
Here $P_0^v$ is the dominant term (Eq.~\ref{PROBVACUUMa}). The CP phase dependence lies in two interference terms, $P^v_{\sin\delta}$ and $P^v_{\cos\delta}$ (see Eq.~\ref{PROBVACUUMb} and \ref{PROBVACUUMc}).
$P^v_{\sin\delta}$ is the CP violating term as it changes sign for antineutrinos, while the $P^v_{\cos\delta}$ is the CP conserving term. It is clear that both $\Delta m_{21}^2$ and $\Delta m_{31}^2$ are required to be non-zero in order to extract any information about the CP phase.   $P^v_3$  is a
 relatively smaller term (Eq.~\ref{PROBVACUUMlast}). 
 
In the presence of standard interactions with matter, Eq.~\ref{probvacuumterms} and Eq.~\ref{PROBVACUUM} are no longer applicable and it is tedious to obtain approximate analytic expressions. It becomes imperative to use certain approximations to derive an analytical expression for $P^m_{\mu e}$. One typically uses constant density approximation and also the fact that $\alpha$ and $\theta_{13}$ are small parameters so that one can employ perturbation techniques to obtain an expression (valid upto second order) for $P^m_{\mu e}$ (see~\cite{Cervera:2000kp,Freund:2001pn,Akhmedov:2004ny}). Here again, we can write down the approximate expression for probability as a sum of four terms which are modified in presence of matter.  
\begin{eqnarray}
P^m_{\mu e}  &\simeq& P^m_0 + P^m_{\sin \delta} + P^m_{\cos \delta} + P^m_3~,
\label{probmatterms}
\end{eqnarray}
where 
\begin{subequations}
\label{probmat}
\begin{eqnarray}
P^m_0 &=&  \sin^2 \theta_{23} \frac{\sin^2 2\theta_{13}}{(\Ap-1)^2} \sin^2((\Ap-1){\Delta_{31}})  ~, \label{probmata}
\\
P^m_{\sin\delta} &=&   \alpha\; \frac{\sin\delta \cos\theta_{13} \sin 2\theta_{12} \sin 2\theta_{13} \sin 2\theta_{23}}{\Ap(1-\Ap)}
 \sin({\Delta_{31}})\sin(\Ap {\Delta_{31}})\sin((1-\Ap) {\Delta_{31}})   ~, \label{probmatb}\\
P^m_{\cos\delta} &=&   \alpha\; \frac{\cos\delta \cos\theta_{13} \sin 2\theta_{12} \sin 2\theta_{13} \sin 2\theta_{23}}{\Ap(1-\Ap)}
 \cos({\Delta_{31}})\sin(\Ap {\Delta_{31}})\sin((1-\Ap) {\Delta_{31}})  ~, \label{probmatc} 
 \\
P^m_3 &=&  \alpha^2 \frac{\cos^2 \theta_{23} \sin^2 2\theta_{12}}{\Ap^2} \sin^2(\Ap {\Delta_{31}})~,\label{probmatd}
\end{eqnarray}
\end{subequations}
where $\Ap = A/\Delta m^2_{31}$ where{{ $A = 2 V E = 2 \sqrt{2} G_F n_e E 
= 2 \times 0.76 \times 10^{-4}  \times Y_e \times [{\rho}/{g \,cm^{-3}}] \times [{E}/{GeV}] ~eV^2 $}}  
  is the Wolfenstein term. $G_F$ is the Fermi  constant and $n_e$ is the electron number density in Earth matter ($n_e = N_{Avo} Y_e \rho$ where, $N_{Avo}$ is 
the Avogadro's number, $Y_e$ is the electron fraction and $\rho$ is Earth matter density). For antineutrinos, $\delta \to -\delta$ and $A \to - A$. {The above expression is valid only when $\alpha \Delta_{31} \lsim 1$ which inturn means that the expression would give reasonable results for
 {{$E > 0.8$ GeV}} for baselines below 8,000 km~\cite{Freund:2001pn}.}

 Note that,  $P_0^m$  is the dominant term as it is independent of $\alpha$ (Eq.~\ref{probmata}). 
We note that the $\delta$-dependent interference terms (Eq.~\ref{probmatb} and \ref{probmatc}) are suppressed by the hierarchy parameter, $\alpha$ and the last term, $P_3^m$ is proportional to $\alpha^2$ (Eq.~\ref{probmatd}).

\subsection{First and second oscillation maxima of $P_{\mu e}$}
\label{max}
Since the leading term depends on $\Delta_{31}$, the physical characteristics of an appearance experiment are therefore
determined by the baseline ($L$) and neutrino energy ($E$) at which the mixing between the $\nu_1$ and $\nu_3$ states
is maximal.  If we look at the leading oscillatory term of $P^v_{\mu e}$, we get  oscillation maxima at 
\begin{eqnarray}
\dfrac{\Delta m^2_{31} L}{4 E} &=& (2n-1) \dfrac{\pi}{2}  \nonumber \\
\dfrac{L}{E} &=& (2n-1) \dfrac{\pi}{2} \left(\dfrac{1}{1.267}\right)   \left(\frac{2.56 \times 10^{-3} ~{\rm{eV^2}} }{\Delta m^2_{31}}\right) 
\nonumber \\ \implies
\dfrac{L}{E} &\simeq & (2n-1) \times {{500}} ~\dfrac{\rm{km}}{\rm{GeV}} 
\end{eqnarray}
\noindent where $n$ is an integer and $n=1,2,\ldots$ stands for first, second, \ldots oscillation maxima occurring at $L/E \simeq
 500, 1500, \ldots$ km/GeV and so on. For a fixed baseline of 1300 km, this would imply $E^{I^{}} 
 \simeq 2.6$ GeV and $E^{II^{}} \simeq 0.87$ GeV  for first and second oscillation maxima respectively.
  It may be possible to observe the higher ($n>1$) oscillation maximas when the 
 baselines are comparatively  longer (so that the energies at which higher maxima  occur are not too small).
 A useful observable to understand the impact of CP phase on the oscillation probabilities (Eqs.~\ref{PROBVACUUM}
and \ref{probmat}) is the CP asymmetry~\cite{Akhmedov:2004ve} and in the next subsection we examine the analytic expression for the CP asymmetry in vacuum and in matter.

\subsection{CP Asymmetry in vacuum and in matter}
\label{acp}

The CP asymmetry is given by
\begin{eqnarray}
A_{\mu e}^{CP}
&=&  
\dfrac{P_{\mu e}(\delta) - \bar{P}_{\mu e}(\delta)}{P_{\mu e}(\delta) + \bar{P}_{\mu e}(\delta)} 
= \dfrac{\Delta P_{\mu e} ^{CP}}{\sum  P_{\mu e}^{CP} }
\end{eqnarray}
Since CP asymmetry is a ratio of probabilities, the systematic uncertainties mostly
 cancel out at the level of event rates.  
In vacuum, the numerator is given by~\footnote{Eq.~\ref{eq:delpv} is obtained by using the exact three flavour formula for 
$P_{\mu e}$ in vacuum given in \cite{Masud:2015xva}.}\begin{eqnarray}
\Delta P^{CP}_{\mu e} &=& 8 J [ \sin (2\alpha\Delta_{31}) \sin^2 (\Delta_{31}) - \sin (2\Delta_{31}) \sin^2 (\alpha\Delta_{31})]~, \nonumber \\
&=& \underbrace{16 \sin\delta J_r [\sin (\Delta_{31}) \sin (\alpha\Delta_{31}) \sin(1- \alpha)\Delta_{31}]}_{\textrm{intrinsic}} ~,
\label{eq:delpv}
\end{eqnarray}
{{where,  
\begin{align}
J_{} &= {\textrm{Im}}\left(U_{\ell' j}^{} U_{\ell j}^\star U_{\ell j'} ^\star U_{\ell' j'} ^ {} \right) \, \quad  {\rm{where}} \quad  l \neq l^\prime \quad {\rm{and}} \quad j \neq j^\prime
 \nonumber \\
 & =  \sin \theta_{12} \cos \theta_{12} \sin \theta_{23} \cos \theta_{23}  \sin {\theta_{13}} \cos^2 \theta_{13} \sin \delta \quad {\rm{using}}  \quad {\cal U} \equiv 
{\cal {U}}_{PMNS}
\label{eq:j}
\end{align} 
is the  Jarlskog factor~\cite{Jarlskog:1985ht,Jarlskog:2004be}. 
As a consequence of the orthogonality of any pair of different rows or columns of the mixing matrix, this imaginary part on right side  of Eq.~\eqref{eq:j} is  unique and can  differ at most by a sign.
}} 
In vacuum, therefore, if the CP phase $\delta$ is either  $0$ or $\pi$, the CP asymmetry vanishes. However, matter effects can create a fake CP asymmetry. 
In case of matter, we have
\begin{eqnarray}
\label{eq:delpm}
\Delta P^{CP}_{\mu e} & \approx & 
\underbrace{\sin^2\theta_{23}  \sin^2 2\theta_{13} \Theta_+\Theta_-}_{\textrm{extrinsic}}  
\nonumber\\ 
&+&
\underbrace{8 \alpha J_r  \frac{\sin (\hat{A}\Delta_{31})}{\hat{A}} [\cos \delta \Theta_{-}  \cos \Delta_{31} +  \sin \delta \Theta_{+}  \sin \Delta_{31}]}_{\textrm{intrinsic and extrinsic}}~,
\end{eqnarray}
where  $\Theta _{\pm} =\sin [(\hat{A} - 1) \Delta_{31}]/(\hat{A}-1) \pm \sin[(\hat{A}+1)\Delta_{31}]/(\hat{A}+1) $. 
The presence of first  term implies that one will have a non-vanishing contribution to
 $\Delta P^{CP}_{\mu e}$ due to the MSW matter effect~\cite{Mikheev:1987qk,Wolfenstein:1977ue} since $\hat A \to - \hat A$ for antineutrinos.   
From Eq.~\ref{eq:delpm}, we note that the $\cos \delta$ dependent term is proportional to $\cos \Delta_{31}$. 
This implies that at location of oscillation maxima 
($\cos \Delta_{31} = 0$), the $\cos \delta$ term vanishes.

\begin{figure}[t!]
\centering
\includegraphics[width=.6\textwidth] {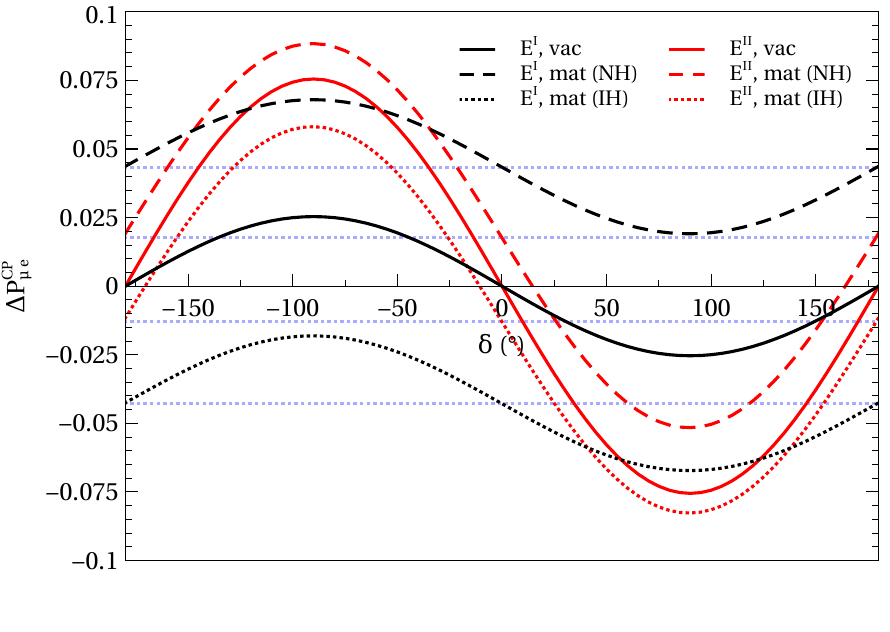}
    \caption{\footnotesize{$\Delta P_{\mu e}^{CP}$ plotted in vacuum (solid lines) and in matter (dashed lines for NH and dotted  lines for IH) as a function of $\delta$ for a fixed baseline of 1300 km. 
    $E^{I} \simeq 2.6$ GeV and $E^{II} \simeq 0.87$ GeV refer to the first (black) and second (red) oscillation maxima respectively.   
  }}
    \label{fig:delpvsdel}
\end{figure}

In Fig.~\ref{fig:delpvsdel}, we plot the probability difference as a function of $\delta$ for a fixed values of energy corresponding to first and second oscillation maxima and baseline of 1300 km (for DUNE). 
The curves are plotted in vacuum and in matter (see Eq.~\ref{eq:delpv} and Eq.~\ref{eq:delpm}). {{It should be noted that all figures are obtained using the General Long Baseline Experiment Simulator (GLoBES) software~\cite{Huber:2004ka, Huber:2007ji} which numerically solves the full three flavour neutrino propagation equations with the Preliminary Reference Earth Model (PREM)~\cite{Dziewonski:1981xy} density profile of the Earth for the values of neutrino parameters listed in Table~\ref{tab:parameters}. The analytic expressions aid in our understanding of the salient features of the considered observables.}} The following comments are in order. 
\begin{itemize}
\item 
From Eq.~\ref{eq:delpv}, we note that $\Delta P^{CP}_{\mu e} \propto \sin \delta$ in vacuum.
While there is a $\cos \delta$ term in presence of matter (Eq.~\ref{eq:delpm}), it vanishes
 (at the location of first and second oscillation maxima as $\cos \Delta_{31} \to 0$) and $\delta$ 
 dependence is via the $\sin \delta$ term in matter as well.
\item 
 $\Delta P^{CP}_{\mu e}$ at $E^{II} > $  $\Delta P^{CP}_{\mu e}$ at $E^{I}$ both in vacuum and in 
matter. The largest difference occurs around $\delta = \pm 90^{\circ}$. At $\delta = - 90^{\circ}$, in vacuum,  
$\Delta P^{CP}_{\mu e} \simeq  0.025$ at  $E^I$ ; $\Delta P^{CP}_{\mu e}  \simeq 0.075 $ at $E^{II}$. 
In matter (for NH), $\Delta P^{CP}_{\mu e} \simeq 0.07 $   at  $E^I$ and  
$\Delta P^{CP}_{\mu e}  \simeq 0.088 $ at $E^{II}$.
\item 
For CP conserving values i.e., $\delta = 0^{\circ}$ or $180^\circ$, in vacuum 
 $\Delta P^{CP}_{\mu e}$ vanishes at  $E^I$ and $E^{II}$. 
But, in matter (for NH), $\Delta P^{CP}_{\mu e} \neq 0 $ due to matter effects (pure extrinsic)
and in fact, $\Delta P^{CP}_{\mu e} \simeq 0.043 $  at  $E^I$ and  
$\Delta P^{CP}_{\mu e}  \simeq 0.017$ at $E^{II}$.
 
\item In general, when $\delta \neq 0$, the extrinsic effects complicate the determination of 
intrinsic CP phase.  As matter effects are more pronounced at $E^{I}$ than at $E^{II}$, the
 intrinsic versus extrinsic separation is harder at $E^{I}$ than at $E^{II}$.

\item The main advantage of studies at second maximum over the first maximum lies in our ability to 
extract the intrinsic CP component better than that with first maximum. This is due to the fact
  that  CP asymmetry in vacuum is larger while the matter effects are not as large. 

 \end{itemize}

\begin{figure}[tbh!]
\centering
\includegraphics[width=.48\textwidth] 
{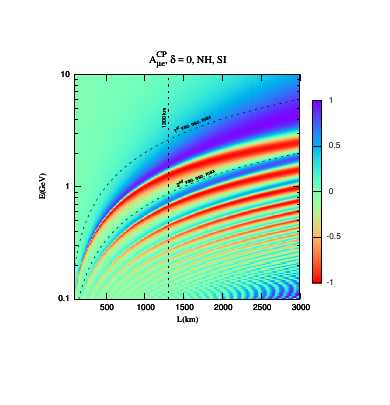}
\includegraphics[width=.48\textwidth] 
{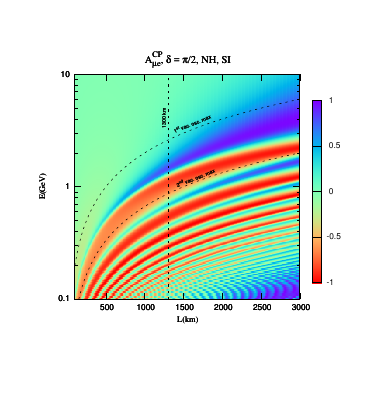}
    \caption{\footnotesize{
 Oscillogram of CP asymmetry in the plane of $E$ and $L$ for $\delta=0$   and $\delta = \pi/2$. 
     }}\label{fig:acposc}
\end{figure}

In order to explore the features of the second oscillation maximum versus the first oscillation maximum as a function of $E$ and $L$, we show oscillograms of CP asymmetry in the plane of $E$ and $L$ in Fig.~\ref{fig:acposc}. The location of first and second oscillation maxima in the plane of $E$ and $L$ is depicted in the plots. {{
For $\delta=0$ (left panel), the CP asymmetry at the second maximum is negligible, as expected. However, matter effects play a crucial role at the first oscillation maximum as can be seen for baselines beyond $L \approx 1000$ km and the  CP asymmetry grows with $L$. For $\delta = \pi/2$ (right panel), one can see significant change in the CP asymmetry at the second oscillation maximum - this is due to the (maximal) intrinsic CP contribution. The pattern of CP asymmetry in the neighbourhood of first maximum is also modified due to the interplay between the intrinsic and extrinsic effects.   The  CP asymmetry has opposite sign (red) at second maximum in comparison to the CP asymmetry at first maximum (blue). 
 This is consistent with Fig.~\ref{fig:delpvsdel}  near $\delta\simeq \pi/2$ for $1300$ km.}}

\subsection{Mass hierarchy asymmetry in vacuum and in matter}
\label{amh}

The MH asymmetry is given by
\begin{eqnarray}
A^{MH}_{\mu e}
&=&  
\dfrac{P_{\mu e} ^{NH} - {P}_{\mu e}^{IH}}{P_{\mu e}^{NH} + {P}_{\mu e}^{IH}} 
= \dfrac{\Delta P^{MH}_{\mu e}}{\sum  P^{MH}_{\mu e}}
\end{eqnarray}
In vacuum,
\begin{equation}
\Delta P_{\mu e}^{MH} \approx 2 J_{r} \sin \D_{31}\sin \D_{21}\cos \D_{31}( \cos \delta \cos \D_{21} - 8\sin \delta \sin \D_{21})
~\label{eq:mhasymm_v}
\end{equation}

\begin{figure}[t!]
\centering
\includegraphics[width=.48\textwidth] 
{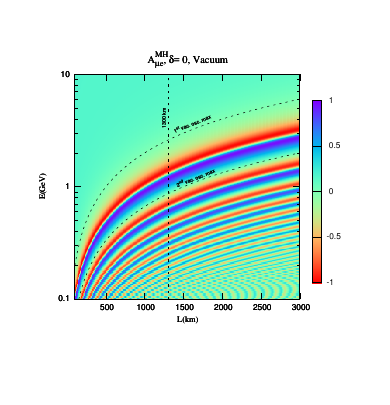}
\includegraphics[width=.48\textwidth] 
{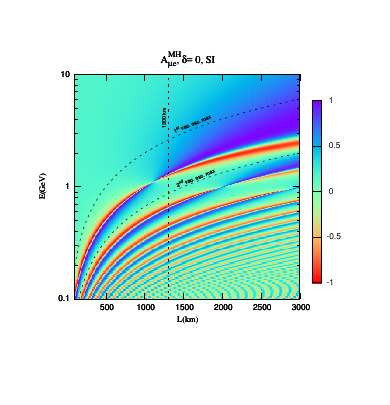}
    \caption{\footnotesize{Oscillogram of MH asymmetry in the plane of $E$ and $L$ for $\delta=0$ in vacuum (left) and matter (right). 
    }}\label{fig:amhosc}
\end{figure}

In matter, 
\begin{eqnarray}
\label{eq:mhasymm_m}
\Delta P^{MH}_{\mu e} \approx \sin^2\theta_{23}  \sin^2 2\theta_{13} \Theta_+\Theta_- + 8 \alpha J_r  \frac{\sin (\hat{A}\Delta_{31})}{\hat{A}} [\cos \delta \Theta_{+}  \cos \Delta_{31} +  \sin \delta \Theta_{-}  \sin \Delta_{31}]~
\end{eqnarray}
where  $\Theta _{\pm} =\sin [(\hat{A} - 1) \Delta_{31}]/(\hat{A}-1) \pm \sin[(\hat{A}+1)\Delta_{31}]/(\hat{A}+1) $ as given before. 


In Fig.~\ref{fig:amhosc}, we show   oscillograms for the MH asymmetry for $\delta=0$ in the plane of $E$ and $L$  for vacuum (left) and matter (right).  
 The  $\Delta P^{MH}_{\mu e} $ is expected to vanish at all nodes (maxima and minima) since $\Delta P^{MH}_{\mu e}  \propto \cos \Delta_{31} \sin \Delta _{31}$ in vacuum (Eq.~\ref{eq:mhasymm_v}). This is independent of the value of $\delta$ (Fig.~\ref{fig:amhosc}, left panel). 
 From the left panel of Fig.~\ref{fig:amhosc}, we note that the MH asymmetry is vanishingly small at both the oscillation maxima.
  If we examine the right panel of Fig.~\ref{fig:amhosc}, 
  at the first oscillation maximum, one expects large matter induced changes in probability in comparison to vacuum which aids the determination of the neutrino mass hierarchy provided the baseline is long enough for matter effects to be substantial.

\subsection{Distinguishing between intrinsic and extrinsic CP contribution}

\begin{figure}[t!]
\centering
\includegraphics[width=.6\textwidth] 
{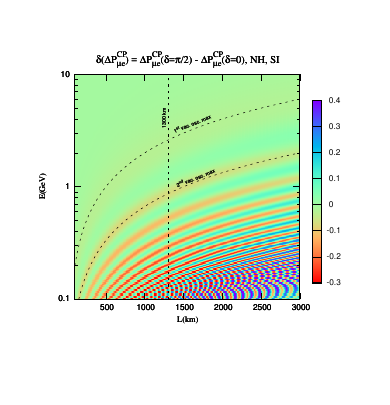}
    \caption{\footnotesize{Oscillogram in the plane of $E$ and $L$ depicting separation of  the intrinsic CP contribution from the extrinsic CP contribution. 
    }}\label{fig:intrinsic}
\end{figure}

In the presence of matter, the CP asymmetry can be expressed in terms of intrinsic and extrinsic CP factors (see Eq.~\ref{eq:delpm}).  
The issue of separation of intrinsic contribution from the extrinsic contribution has been addressed in~\cite{Nunokawa:2007qh,Rout:2017udo} and a useful observable (matter contribution gets approximately cancelled) to disentangle the two contributions was suggested. The observable is given by 
\begin{eqnarray}
\delta(\Delta P^{CP}_{\mu e}) &=& \Delta P^{CP}_{\mu e} (\delta = \pi/2) - 
\Delta P^{CP}_{\mu e} (\delta =0)~
\end{eqnarray}
Using this observable, we next depict an oscillogram in the plane of $E$ and $L$ to illustrate separation of the intrinsic versus exrinsic CP contribution in Fig.~\ref{fig:intrinsic}.  
It is clear that the second oscillation maximum allows for a clean extraction of the intrinsic CP phase. 
This can be easily explained by Fig.~\ref{fig:delpvsdel}, from which we noted that matter influences the probability at first oscillation maximum far more than it impacts the probability at the location of second oscillation maximum. This point and the fact that the intrinsic CP asymmetry is larger at second oscillation maximum makes the second oscillation maximum the preferred option for addressing the question of separating the intrinsic contribution to the CP phase. 

A closer examination of the features in both Fig.~\ref{fig:acposc} and ~\ref{fig:amhosc} reveal a key phenomenon that is often overlooked in studies of the CP asymmetries in long-baseline oscillations. Both the intrinsic and extrinsic CP asymmetries are largest and changing the most rapidly not at the oscillation maxima but in the region between the maxima - near the oscillation minima. Experimentally, there are few events appearing near oscillation minima, so long-baseline experiments are typically designed to maximize the flux of neutrinos in the region right at the first oscillation maxima - which also occurs at the highest energies and therefore highest interaction cross-sections and appearance statistics. The optimization of the DUNE experiment specifically targeted optimizing the beam flux to fully cover the region of the $1^{st}$ oscillation maxima as well as the region  in between $E^{I}$ and $E^{II}$ using a genetic algorthim (GA) to optimize the wide-band beam design to maximize the sensitivity to CP violation.  {{This optimized beam, with a decay pipe 195 m long and 4 m in diameter, produces a muon neutrino flux that is 20\% greater than the nominal configuration at the first oscillation maximum (between 1.5 and 4 GeV), 53\% greater at the second oscillation maximum (between 0.5 and 1.5 GeV), and reduces the antineutrino contamination of the beam.}} As shown in the DUNE Conceptual Design Report~\cite{Acciarri:2015uup}, the GA optimization increased the flux primarily in the region between $E^I$ and $E^{II}$ and {{significantly reduced the flux in the region above the $E^{I}$ oscillation node which contributes to backgrounds below $E^I$. }}  It is for this reason that DUNE can reach the same sensitivities to the 3-flavour oscillation parameters using smaller mass detectors compared to similar experiments with flux only in the region around $E^{I}$.

Our discussion, thus far, has been  at the level of probabilities and in order to 
 obtain realistic quantitative results, one needs to check the outcome at the level of event rates. 
 The following section covers the experimental details used in our simulations.

\section{Experimental inputs}
\label{expt}

\subsection{Beam tunes used}
\label{beamtunes}

\begin{figure}[htb!]
\centering
\includegraphics[width=0.6\textwidth]{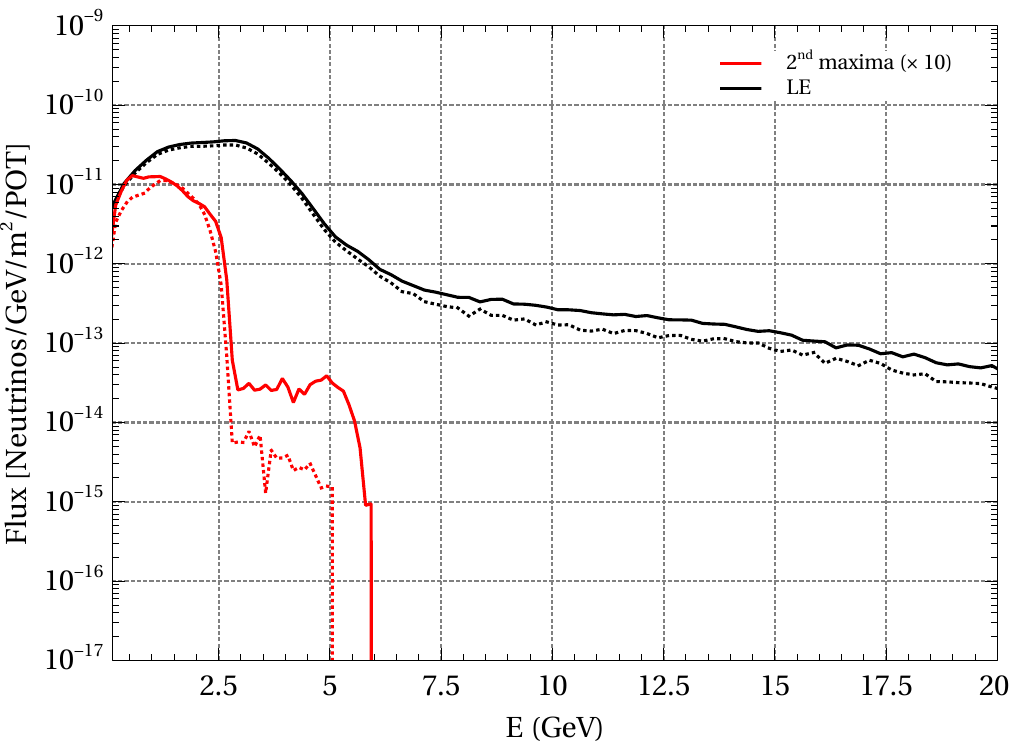}
\caption{\footnotesize{
{{The  fluxes used in the present study. The LE flux corresponds to the CP optimized 80 GeV beam (as discussed in Conceptual Design Report (CDR)~\cite{Acciarri:2015uup}) and the $2^{nd}$ maxima flux is obtained from a Monte Carlo simulation of an 8 GeV beam using the same target and focusing system design as for the CDR 80 GeV optimized beam. The solid line corresponds to the neutrino mode while the dotted line corresponds to the anti-neutrino mode. The fluxes are given in neutrinos/GeV/m$^2$/proton-on-target (POT). Note that for equal average beam power, the number of total delivered POT for the 8 GeV beam will be 10 times larger than that of the 80 GeV beam. 
}
}
}}
\label{fig:flux}
\end{figure}

    The fluxes considered in the present work are plotted as a function of energy in 
Fig.~\ref{fig:flux}.

\begin{itemize}
\item 
{\bf{LE flux :}}
The standard wide-band beam  used in our studies of DUNE is derived from a proton beam of energy  80 GeV~\cite{Acciarri:2015uup,dunefluxes} coupled with a target and focusing system design optimized using a genetic algorithm (GA) for maximal sensitivity to CP violation.  By virtue of its wide-band nature, the LE beam is  also sensitive to regions above and below the first oscillation maximum and covers part of the region of the second oscillation maximum. In the initial running period of DUNE - here assumed to be 5 years, the LE beam will operate at 1.1MW of power from the Fermilab accelerator complex after the Proton Improvement Plan (PIP) II upgrades. A further set of upgrades aimed at replacing the aging 8 GeV booster (PIP-III) ~\cite{pip} is planned during DUNE operations which will double the power of the LE beam to 2.2 MW @ 80 GeV. 
  \item {\bf{$2^{nd}$ maxima flux :}} The neutrino beam at the $2^{nd}$ maxima is obtained using a 3 MW, 8 GeV proton beam which could be generated by the PIP-III SRF linac option~\cite{pip}. The target and focusing system design is assumed to be identical to the LE beam system. If PIP-III is realized using the pulsed SRF linac option, both the 2.2 MW 80 GeV wide-band beam and the 8 GeV 3MW beam could potentially run simultaneously which is what is assumed in this study. 
  \end{itemize}

 The beamline details for the two fluxes are listed in Table~\ref{tab:flux}.
\begin{table}[htb!]
\centering
{{
\begin{tabular}{ l  l  l }
\hline
&&
\\
Parameter & LE (CPV optimized design) &$2^{nd}$ maxima    \\
&&
\\
\hline 
Proton beam energy & 80 GeV & 8 GeV \\ 
Proton Beam power & 1.1 MW (PIP-II)/2.2 MW (PIP-III) & 3 MW (PIP-III)\\
Protons on target (POT) per year  & $1.47 \times 10^{21}$/$2.94 \times 10^{21}$  & $40.1 \times 10^{21}$ \\
Focusing &  \multicolumn{2}{c}{2 horns, GA optimized for CPV sensitivity (2015)} \\
 Horn Current  & $\sim$ 300 kA & $\sim$ 300 kA \\
Decay pipe length & 194 m & 200 m \\
Decay pipe diameter & 4 m &  4 m \\
 && \\
& GA $\to$ Genetic Algorithm & \\
\hline
\end{tabular}
}}
\caption{\label{tab:flux} {\footnotesize{
Beamline parameters assumed for the different design fluxes used  in our sensitivity calculations~\cite{2013arXiv1307.7335L, Alion:2016uaj, Bishai:2013yqo}. 
{The LBNF neutrino beamline decay pipe length has been chosen to be 194 m. }}}
} 
\end{table}

%
\subsection{Detector details}
\label{detdetail}

{{The DUNE far detector (FD) is described in detail in Volume 4~\cite{Vol4:CDR} of the DUNE CDR. }}
 In our analysis, we have assumed that a Liquid Argon (LAr) far detector of fiducial mass $40$ kt situated at a distance of $1300$ km.
 We have combined both electron neutrino appearance ($\nu_\mu \to \nu_e$) and muon neutrino disappearance ($\nu_\mu \to \nu_\mu$) channels, both in neutrino and anti-neutrino mode. Details of the  experimental configuration and other parameters relevant 
 for DUNE are given in Table~\ref{tab:sys}.

%
\begin{table}[hbt!]
\centering
\begin{tabular}{ |l| l | l | }
\hline
{{S. No.}} & {{Relevant parameter}}  & {{DUNE}} \\
\hline
&& \\                         
1 & Location & USA \\
2 & Status & under construction \\
3 & Accelerator facility & Fermilab \\
4 & Beam power & $1.1$ MW @ 80 GeV (PIP-II), $2.2$ MW @ 80 GeV (PIP-III)\\
& & $3$ MW @ 8 GeV (PIP-III SRF linac option)\\
5 & Expected POT/yr & See Table~\ref{tab:flux} \\
6 & Baseline length & 1300 km \\
7 & Off-axis angle & $0^\circ$ \\
8 &  Detector technology & LAr \\
9 & Fiducial mass  & 40 kton  \\
10 & Run times & 5 yr with 80 GeV @ 1.1 MW \\
& & + 5 yr with 80 GeV @ 2.2 MW \\
& & + 5 yr with 8 GeV @ 3 MW\\
11 & Energy window (GeV) & [0.125,18.0] \\
12 & Energy bins & 71   \\
13 & Energy resolution, $\sigma_e$ & migration matrices based on fast MC (2015 CDR)\\
14 & Normalization error  & $\nu_e : 2\%$ (signal)  $\nu_e : 5\%$ (bkgd) \\ 
 &  & $\nu_\mu : 5\%$ (signal)  $\nu_\mu : 5\%$ (bkgd) \\                                
\hline
\end{tabular}
\caption{\label{tab:sys} {\footnotesize{
Details of the  experimental configuration and other parameters relevant for DUNE.  }}}
\end{table}

\section{Event spectrum at DUNE}
\label{event}

In order to simulate DUNE, we use the GLoBES package~\cite{Huber:2004ka, Huber:2007ji} with the DUNE configuration file provided by the collaboration~\cite{Alion:2016uaj} {{as well as for a scenario with improved energy reconstruction capabilities}}.    
We implement the density profile of Earth as  given by PREM~\cite{Dziewonski:1981xy}.  
 {{We obtain our results for  the following two scenarios : 
\begin{enumerate}
\item[(i)] 
the standard case using the available DUNE configuration files~\cite{Alion:2016uaj} with the signal and background smearing matrices obtained from a fast MC (2015 CDR), and 
\item[(ii)] a case corresponding to improved energy reconstruction capabilities; we implement this by introducing an energy dependent Gaussian smearing in GLoBES with a resolution of $\Delta E/E = 10\%/\sqrt{E {\rm (GeV)}}$ and 
  $15\%/\sqrt{E {\rm (GeV)}}$ respectively for the  $\nu_e$ CC and $\nu_\mu$ CC signal  and backgrounds, The NC and $\nu_\tau$ backgrounds remain the same as in (i). We note that in the latest
  studies of DUNE performance reported in the 2020 Technical Design Report, the average energy resolution of $\nu_e$ and $\nu_\mu$ CC events obtained using full simulation and reconstruction is 13\% and 18\% respectively~\cite{Abi:2020evt}. It is possible that further improvements in reconstruction algorithms could reach the performance we assumed above.
\end{enumerate}}}

We consider the following beam and runtime combinations :
\begin{itemize}
\item[(a)]  
 LE, 1.1 MW + LE, 2.2 MW $\to$ runtime of 5 yr in LE, 1.1 MW and 5 yr in LE, 2.2 MW
  distributed equally in  neutrino and antineutrino modes
 \item[(b)]  
  {{$2^{nd}$ maxima, 3 MW}} $\to$  runtime of 5 yr distributed equally in  neutrino and antineutrino modes
 \item[(c)]   {{LE, 1.1 MW}} + {{LE, 2.2 MW}} +
  {{$2^{nd}$ maxima, 3 MW}} $\to$  runtime of 5 yr in LE, 1.1 MW,  
   5 yr in LE, 2.2 MW and 5 yr with $2^{nd}$ maxima, 3 MW 
   distributed equally in neutrino and antineutrino modes. It is assumed the 8 GeV 3MW beam option ($2^{nd}$ maxima) runs concurrently with the 80 GeV 2.2 MW (LE) default beam using the same focusing system. 
  \end{itemize}  
  Table~\ref{tab:events} lists the total signal and background 
  events for the three beam combinations mentioned above for the $\nu_e$ appearance and  
  $\nu_\mu$ disappearance channels.  
{{In Table~5, we list the number of signal and total background events in the region of the secondary oscillation nodes ($2^{nd}$ maxima region and below) in the two beam options from the Fermilab PIP-III upgrade - the 2.2 MW 80 GeV beam and the 3 MW 8 GeV beam. The addition of running with the 8 GeV 3MW enhances the default DUNE event yield at the secondary oscillation nodes by $\sim 50-60\%$.}}

\begin{figure}[t!]
\centering
\includegraphics[width=0.85\textwidth]{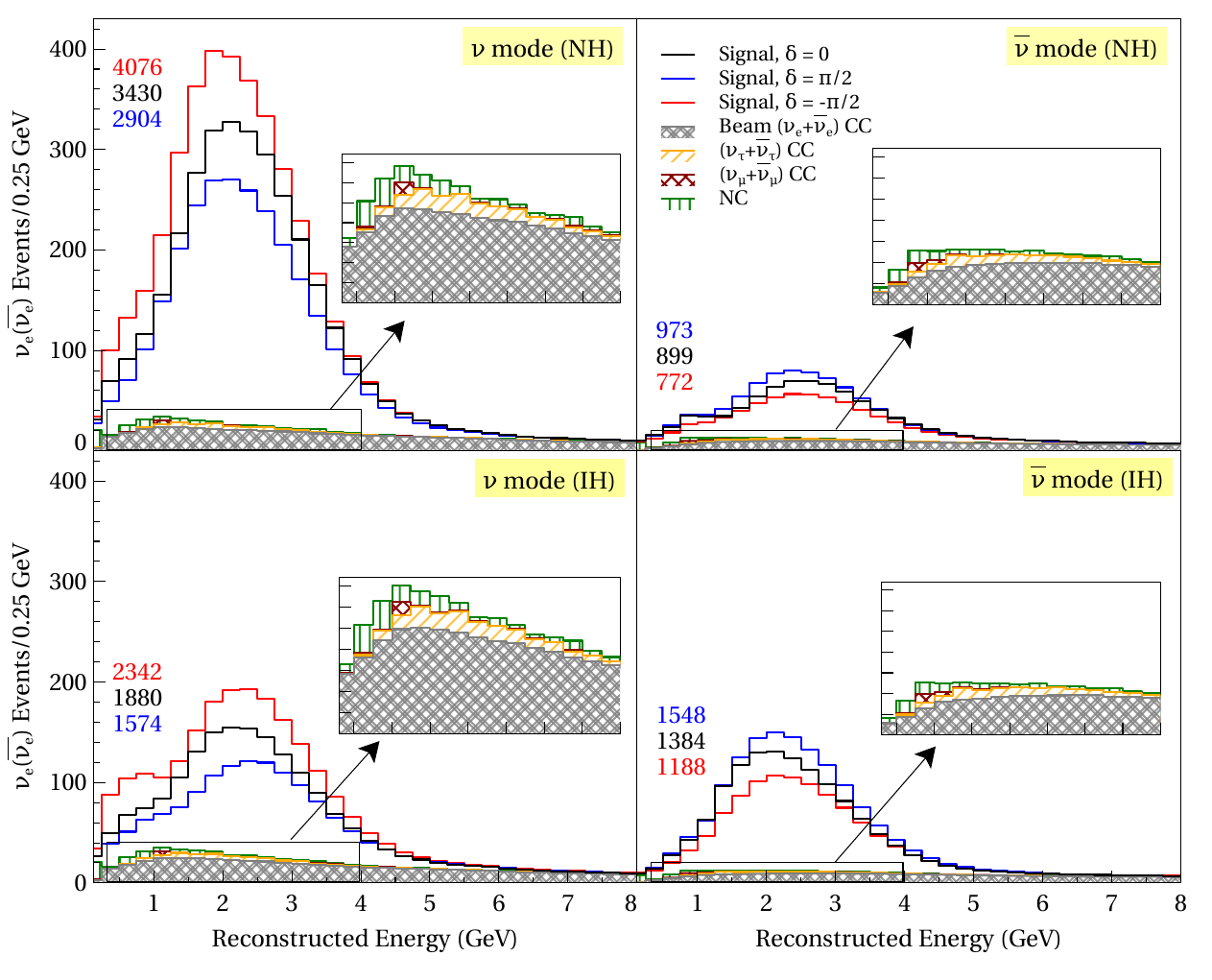}
\caption{\footnotesize{Electron (anti-)neutrino appearance event spectrum with a 40 kt DUNE far detector {{using DUNE configuration files~\cite{Alion:2016uaj}}} for the beam combination (c) : LE 1.1 MW beam,  LE 2.2 MW beam and $2^{nd}$ maxima 3 MW beam for different values of $\delta$ ($\delta=0, \pi/2, -\pi/2$). 
The runtime of 15 yr is split equally among the three beams (5 yr each) and distributed evenly between neutrino and  antineutrino modes (2.5 yr +2.5 yr) for each beam tune.
{{
 The event rates are stacked from below backgrounds. The total number of events corresponding to a particular value of $\delta$ are mentioned in each panel. }}  
  }} 
\label{fig:e3}
\end{figure}

\begin{figure}[t!]
\centering
\includegraphics[width=0.85\textwidth]{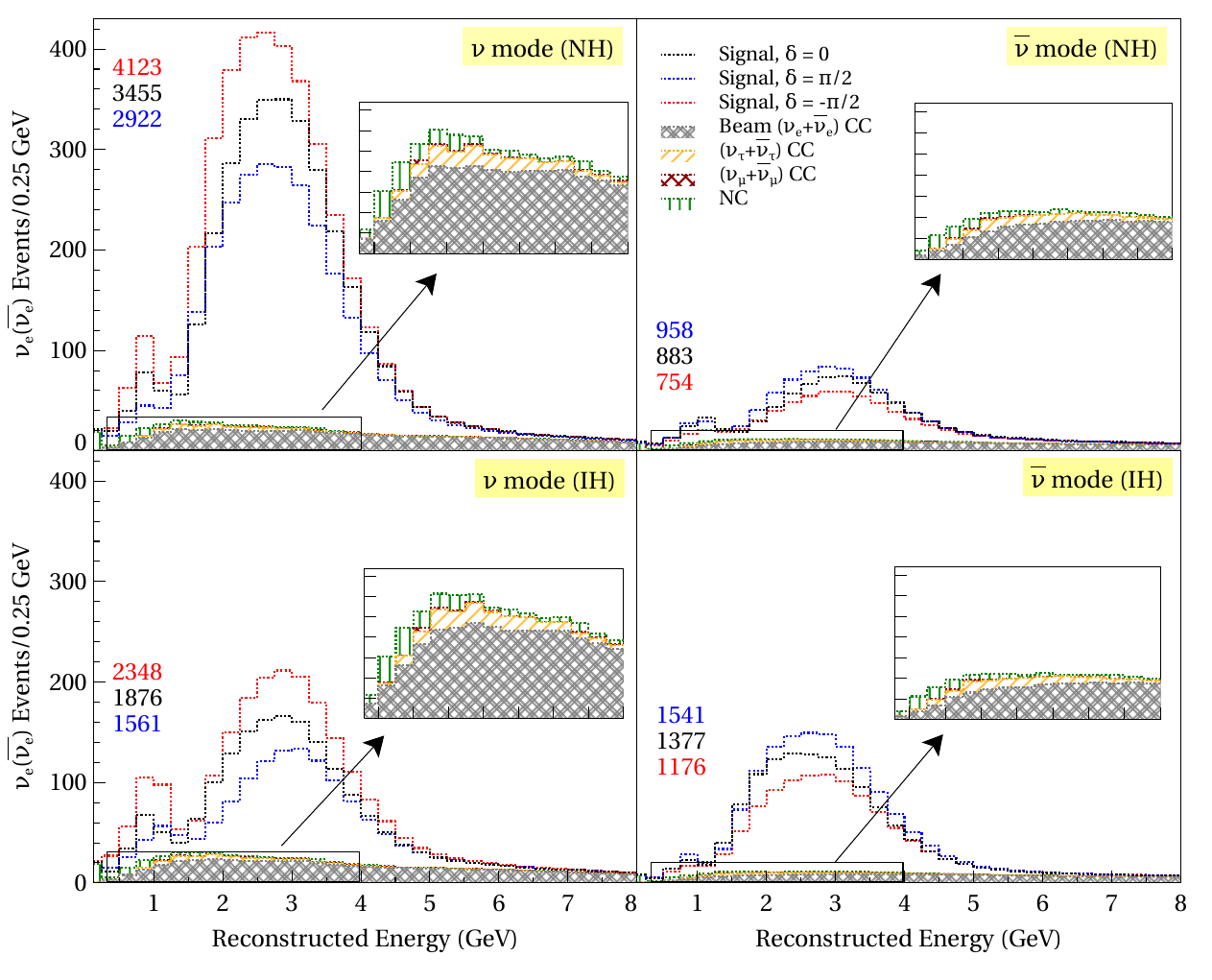}
\caption{\footnotesize{{{Same as Fig.~\ref{fig:e3} with improved energy reconstruction capabilities shown using dotted lines. }}}}
\label{fig:e3g}
\end{figure}

\begin{figure}[hbt!]
\centering
\includegraphics[width=0.85\textwidth]{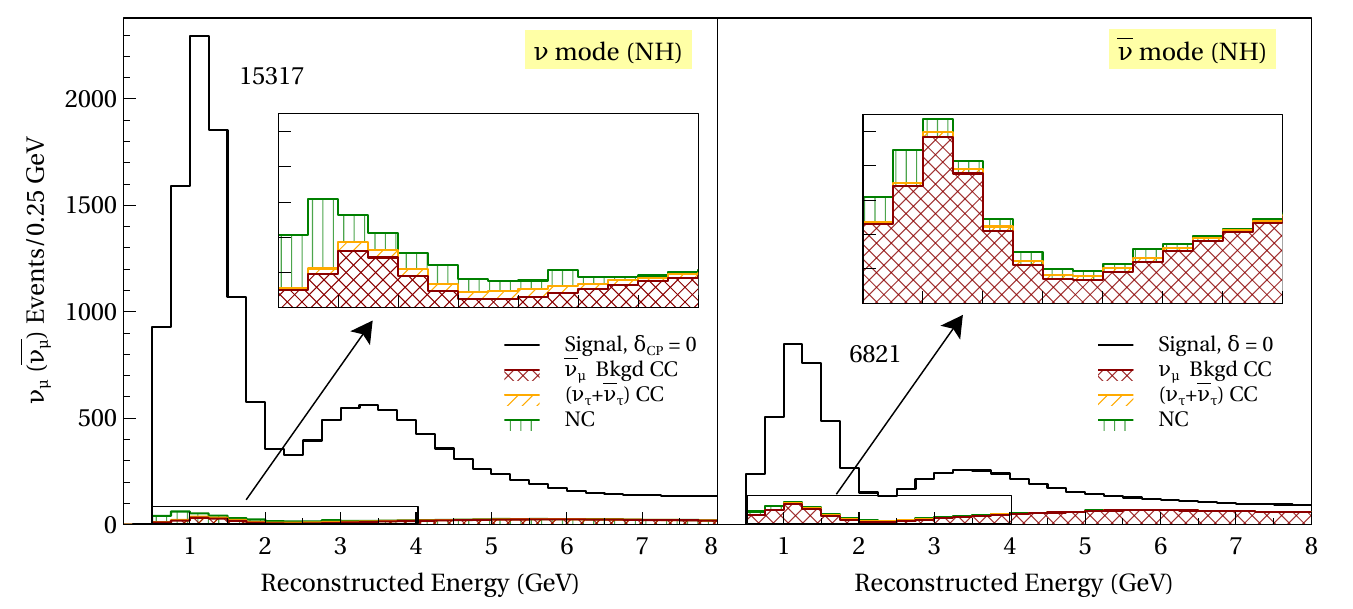}
\caption{\footnotesize{Muon (anti-)neutrino disappearance event spectrum with a 40 kt DUNE far detector for the combination (c) : LE 1.1 MW beam,  LE 2.2 MW beam and $2^{nd}$ maxima 3 MW beam for $\delta =0$. 
The runtime is split equally among the three beams (5 yr each) and distributed evenly between neutrino and  antineutrino modes (2.5 yr +2.5 yr).  
  }} 
\label{fig:e4}
\end{figure}

\begin{figure}[hbt!]
\centering
\includegraphics[width=0.85\textwidth]{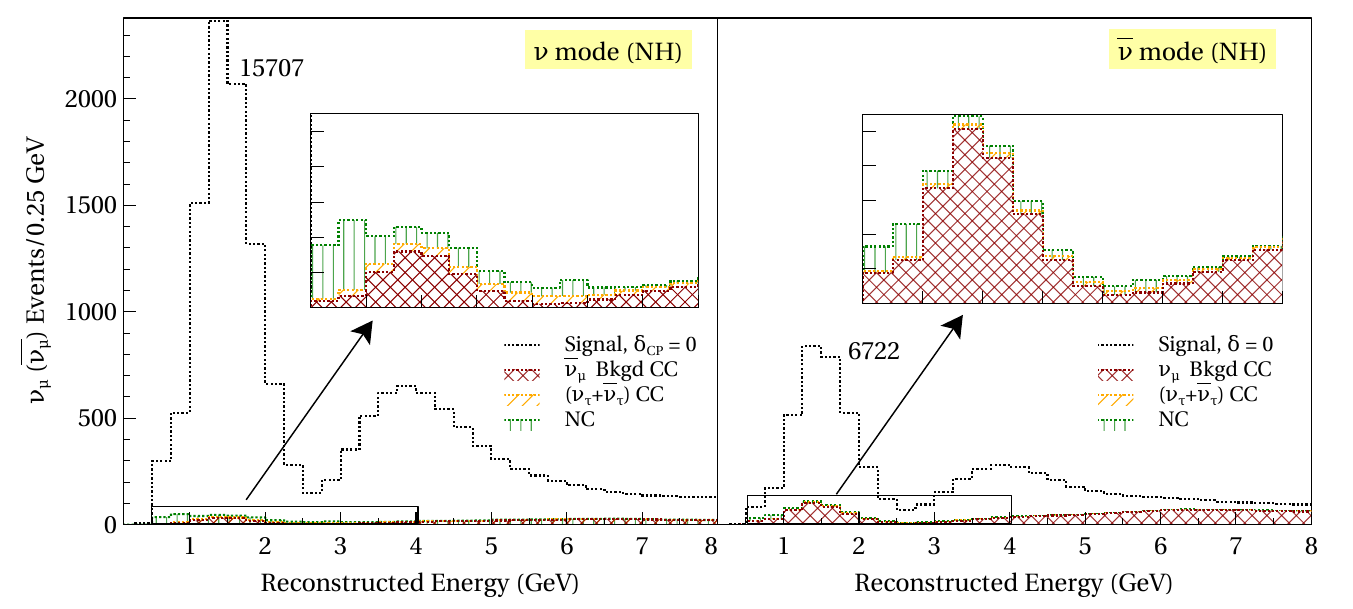}
\caption{\footnotesize{{{Same as Fig.~\ref{fig:e4} with improved energy reconstruction capabilities shown using dotted lines.}}
  }} 
\label{fig:e4g}
\end{figure}

   The event spectrum for option (c) is shown {{in  Figs.~\ref{fig:e3}  - \ref{fig:e4g}}}. 
{{Fig.~\ref{fig:e3}  and Fig.~\ref{fig:e3g} correspond to $\nu_e$ appearance  events  while Fig.~\ref{fig:e4} and Fig.~\ref{fig:e4g}
    show the $\nu_\mu$ disappearance events for the two scenarios.}}  In general, one notes that the events are peaked at the value of energy where the flux is largest - for 80 GeV LE beam, the events peak around $2-3$ GeV while for $8$ GeV beam, it is around $0.8-0.9$ GeV. The $\delta$-dependence of the event spectrum  can be understood from the $\delta$ dependence of the probabilities.


\begin{table}[ht!]
{
\scriptsize
{
\centering
\begin{tabular}{|l | c c | c c | c  | c  |}
\hline
 {\bf{Beam options}}   & \multicolumn{2}{c |}{$\nu_\mu \to \nu_e$} & \multicolumn{2}{c |}{$\bar\nu_\mu \to \bar\nu_e$} & $\nu_\mu \to \nu_\mu$ & $\bar\nu_\mu \to \bar\nu_\mu$ \\

\cline{2-7} & NH & IH & NH & IH & NH  & NH \\
\hline
{\bf{(a) LE, 1.1 MW + LE, 2.2 MW}} &&&&&&\\
Signal $\delta= 0$                  & 3222 & 1759 & 859 & 1303 & 13415 &  6158 \\
Signal $\delta= \pi/2$              & 2727 & 1481 & 928 & 1463 &       &       \\
Signal $\delta= -\pi/2$             & 3784 & 2167 & 742 & 1130 &       &       \\
Bkgd ($\bar\nu_{e}+\nu_e$) CC       & 446  & 461  & 227 &  224 &       &       \\
Bkgd ($\bar\nu_{\mu}+\nu_\mu$) CC   &  6   &  6   &  4  &   3  &       &       \\
Bkgd ($\bar\nu_{\tau}+\nu_\tau$) CC & 43   & 44   & 25  &  25  &   61  &   38  \\
Bkgd NC                             & 55   & 55   & 27  &  27  &  155  &   81  \\
Bkgd $\nu_\mu$ CC                   &      &      &     &      &       &  1536 \\
Bkgd $\bar\nu_{\mu}$ CC             &      &      &     &      &  547 &       \\
 \hline
{\bf{(b) $2^{nd}$ maxima, 3 MW}} &&&&&&\\
Signal $\delta= 0$                  & 208 & 121 & 40 & 82 &  1902 &  663 \\
Signal $\delta= \pi/2$              & 178 & 93  & 45 & 85 &       &       \\
Signal $\delta= -\pi/2$             & 292 & 175 & 30 & 58 &       &       \\
Bkgd ($\bar\nu_{e}+\nu_e$) CC       & 37  & 38  & 13 & 12 &       &       \\
Bkgd ($\bar\nu_{\mu}+\nu_\mu$) CC   &  1  &  1  & 1  & 1  &       &       \\
Bkgd ($\bar\nu_{\tau}+\nu_\tau$) CC &  0  &  0  & 0  & 0  &   0   &   0  \\
Bkgd NC                             &  4  &  4  & 1  & 1  &  11   &   5  \\
Bkgd $\nu_\mu$ CC                   &     &     &     &   &       &  27 \\
Bkgd $\bar\nu_{\mu}$ CC             &     &     &     &   &  8    &       \\
 \hline
{\bf{(c) LE, 1.1 MW + LE, 2.2 MW}}  &&&&&& \\
+ {\bf{$2^{nd}$}} 
{\bf{maxima, 3 MW}} &&&&&&\\
Signal $\delta= 0$                  & 3430 (3455) & 1880 (1876) & 899 (883)& 1384 (1377) & 15317 (15707)&  6821 (6722) \\
Signal $\delta= \pi/2$              & 2904 (2922) & 1574 (1561) & 973 (958) & 1548 (1541) &       &       \\
Signal $\delta= -\pi/2$             & 4076 (4123) & 2342 (2348) & 772 (754) & 1188 (1176) &       &       \\
Bkgd ($\bar\nu_{e}+\nu_e$) CC       & 483  & 499  & 239 &  236 &       &       \\
Bkgd ($\bar\nu_{\mu}+\nu_\mu$) CC   &  7   &  7   &  4  &   4  &       &       \\
Bkgd ($\bar\nu_{\tau}+\nu_\tau$) CC & 43   & 44   & 25  &  25  &   61  &   38  \\
Bkgd NC                             & 59   & 59   & 28  &  28  &  166  &   86  \\
Bkgd $\nu_\mu$ CC                   &      &      &     &      &       &  1563 \\
Bkgd $\bar\nu_{\mu}$ CC             &      &      &     &      &  555  &       \\
\hline
\end{tabular}
\caption{\footnotesize{Total number of event rates for different  beamtune combinations (the signal events for combination (c) when improved energy reconstruction is taken into account are given in brackets).}}
}}
\label{tab:events}
\end{table}


\begin{table}[hbt!]
\label{events2}
{
\scriptsize
{\centering
\begin{tabular}{|l | c | c   |}
\hline
 {\bf{Beam options}}   & \multicolumn{2}{c |}{$\nu_\mu \to \nu_e$} 
  \\
\cline{2-3} & NH & IH    \\
\hline
{\bf{(a)  LE, 2.2 MW}} &&\\
Signal $\delta= 0$                  & 373 (134) & 219 (111)   \\
Signal $\delta= \pi/2$              & 333 (115) & 175 (105)        \\
Signal $\delta= -\pi/2$             & 508 (181) & 290 (169)        \\
Bkgd         & 88 (65)  & 90 (65)      \\
 \hline
{\bf{(b) $2^{nd}$ maxima, 3 MW}} &&\\
Signal $\delta= 0$                  & 120 (75)  & 77 (58)   \\
Signal $\delta= \pi/2$              & 100 (55) &  62 (51)      \\
Signal $\delta= -\pi/2$             & 177 (110) & 125 (107)      \\
Bkgd       &  27 (21)  & 28 (21)        \\
 \hline
\end{tabular}
\caption{\footnotesize{Number of $\nu_\mu \to \nu_e$ CC events and total 
 background in the region of the secondary oscillation nodes  (0.125 - 1.5 GeV {\it{reconstructed}} energy) from
   the PIP-III 80 GeV 2.2 MW and 8 GeV 3 MW beams (we have listed the events for improved energy reconstruction in brackets).}}
}}
\end{table}

The signal for $\nu_e$ appearance is an excess of charged-current (CC) $\nu_e$ and $\bar\nu_e$ interactions over the
expected background in the far detector. {{The background to $\nu_e$ appearance is composed of
\begin{itemize}
\item Beam  $(\nu_e + \bar\nu_e)$ CC : CC interactions of $\nu_e$ and $\bar\nu_e$ intrinsic to the beam; 
\item 
$ (\nu_\tau + \bar\nu_\tau)$ CC :    $\nu_\tau$ and $\bar\nu_\tau$ CC events in which the $\tau$'s decay leptonically into electrons/positrons.
  \item  $(\nu_\mu + \bar \nu_\mu) $ CC :   misidentified $\nu_\mu$ and $\bar\nu_\mu$ CC events; 
\item NC : neutral current 
backgrounds 
\end{itemize}
It should be noted that though NC and $\nu_\tau$ backgrounds are due to interactions of higher-energy neutrinos, they contribute to backgrounds mainly at
lower energies which impacts  sensitivity to CP violation.}}

  \section{Analysis procedure}
  \label{analysis}
To estimate the sensitivities of DUNE to CP violation,   MH,   and octant of $\tc$, we perform a standard $\chisq$ analysis. Even though all results are produced numerically with the help of the GLoBES software, in order to gain insight, let us  examine the analytical form of the $\chisq$ relevant for each of the mentioned 
unknowns. 
 

  {\bf{Sensitivity to CP violation :-}}
Including only statistical effects, the $\chi^2$ for CP violation sensitivity for a given oscillation channel (say $\nu_\mu \to \nu_e $) is given by~\cite{Masud:2015xva,Masud:2016bvp}
\begin{equation}
\label{chisq_cp}
\chi^2 \equiv  \min_{\delta_{test}}  \sum_{i=1}^{x}  \sum_{j=\nu, \bar\nu}^{} 
 \frac{\left[N_{true}^{i,j}(\delta_{true}) - N_{test}^{i,j} (\delta_{test}=0,\pi )\right]^2 }{N_{true}^{i,j} (\delta_{true})}~
\end{equation}
where $N_{true}^{i,j}$ and $N_{test}^{i,j}$ are the number of true and test events in the $\{i,j\}$-th 
bin respectively~\footnote{$N_\sigma = \sqrt{\Delta \chi^2}$. 
$\Delta \chi^2 = \chi^2$ as we have not included any fluctuations 
in simulated data. 
This is the Pearson's definition of $\chi^2$~\cite{Qian:2012zn}. For large sample size, the other 
 definition using log-likelihood also yields similar results.}. 
 The index  $i$ corresponds to energy bins  ($i=1 \to x$, the number of bins depends upon the particular experiment under consideration. For the case of
 \dune, there are 64 
 bins of width 125 MeV in $0-8$ GeV and 7 unequal bins in $8-20 $ GeV. $j$ is being summed over the neutrino and anti-neutrino contribution.  In order to determine the $ \chi^2$ that concerns the  sensitivity to CP violation, the test value of phase ($\delta$) is assumed to be $0$ or $\pi$ and
  the  $ \chi^2$  for any true value of phase ($\delta$) in the full range of $[-\pi,\pi]$ is computed.

The characteristic double peak shape of the curves are expected since the sensitivity drops to zero at the CP conserving values while it is maximum at the maximal CP violating values ($\delta = \pm \pi/2$). 
  The $\chi^2$ is computed as given in Eq.~\ref{chisq_cp} for a given set of true values by minimizing over the test parameters and this procedure is repeated for all possible true values  listed in Table~\ref{tab:parameters}. 
Some general remarks relevant for computation of $\chi^2$ for this and other unknowns are in order. We have marginalised over the standard oscillation parameters.  It should be noted that the total $\chi^2$ is a sum of contributions from the two channels ($\nu_\mu \to \nu_e$ and $\nu_\mu \to \nu_\mu$).   \\

    {\bf{Sensitivity to the MH :-}} The determination of the MH is a measurement of a binary value, namely the sign of the difference of the square of the masses of the 1 and 3 states, Therefore, there can only be two possibilities for the
 true choice of hierarchy (NH or  IH) and one would like to decipher the sensitivity of DUNE to 
figure out the correct MH.

In order to understand the features of the sensitivity plots (considering the true hierarchy as NH or IH), 
we give a  statistical definition of $\chi^2$ as follows~\cite{Masud:2016gcl}.
\begin{equation}
\label{chisq_mh}
\chi^2_{} \equiv  \min_{\delta_{test}}  \sum_{i=1}^{x}  \sum_{j=\nu,\bar \nu }^{} 
 \frac{\left[N_{NH}^{i,j}(\delta_{true}) - 
 N_{IH}^{i,j} (\delta_{test}  )\right]^2 }
 {N_{NH}^{i,j} (\delta_{true})}~
\end{equation}
where the true hierarchy is taken to be NH. $N_{NH}^{i,j}$ and $N_{IH}^{i,j}$ are the number of NH and IH events in the $\{i,j\}$-th bin respectively. %
 The index 
 $i$ corresponds to energy bins  ($i=1 \to x$, the number of bins depends upon the particular experiment under consideration. 
 For the case of \dune, there are 64 bins of width 125 MeV in $0-8$ GeV and 7 unequal bins in $8-20 $ GeV. $j$ is being summed over the neutrino and anti-neutrino contribution.

\begin{figure}[hbt!]
\centering
\includegraphics[width=.8\textwidth]{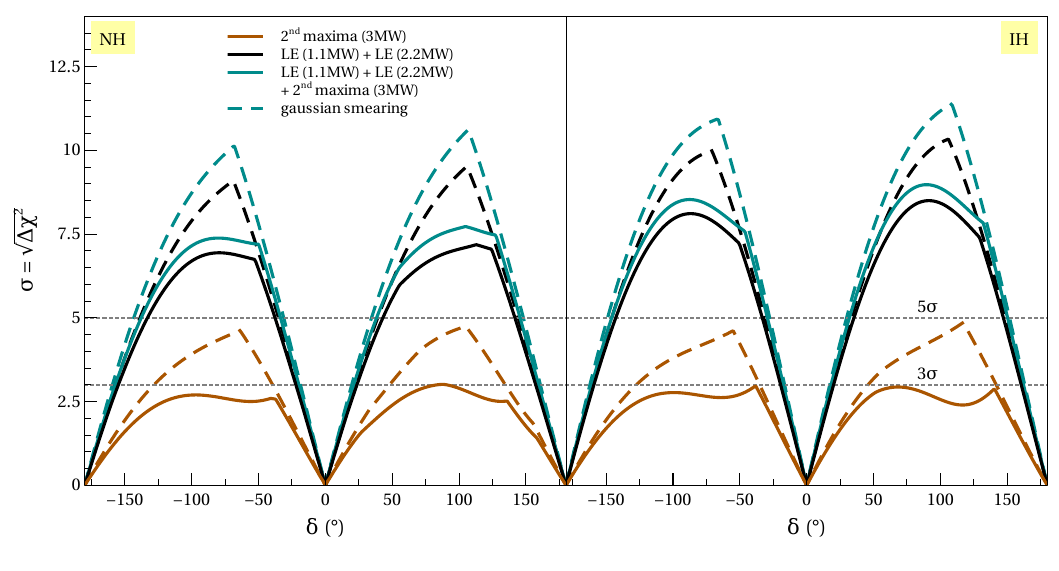}
\caption{\footnotesize{Sensitivity to CP violation as a function of $\delta$. The three curves correspond to (a) : $2^{nd}$ maxima 3MW beam  (2.5yr $\nu$+ 2.5yr $\bar\nu$), (b) : LE 1.1 MW beam (2.5yr $\nu$+ 2.5yr $\bar\nu$) + LE 2.2 MW beam (2.5yr $\nu$+ 2.5yr $\bar\nu$) and (c) :  LE 1.1 MW beam  (2.5yr $\nu$+ 2.5yr $\bar\nu$) + LE 2.2 MW beam  (2.5yr $\nu$+ 2.5yr $\bar\nu$) + $2^{nd}$ maxima 3MW beam  (2.5yr $\nu$+ 2.5yr $\bar\nu$). 
{{The solid lines (in this and the following figures) depict the standard case with the DUNE configuration files~\cite{Alion:2016uaj} 
 while the dashed lines depict the scenario with improved neutrino energy resolution (Gaussian smearing). 
 }}
}}
\label{fig:9m}
\end{figure}
\begin{figure}[hbt!]
\centering
\includegraphics[width=.8\textwidth] 
{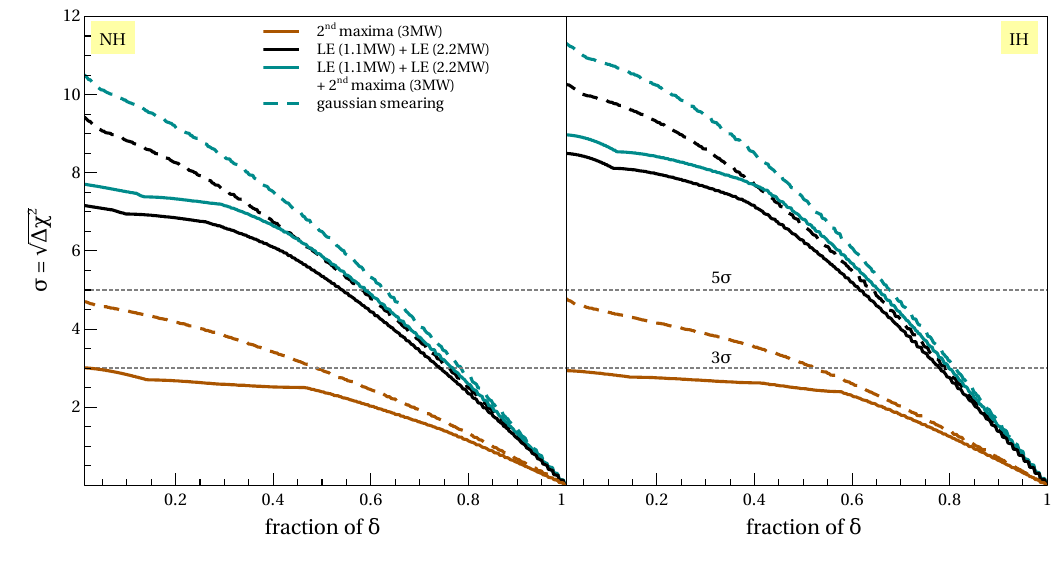}
\caption{\footnotesize{Sensitivity to CP violation  as a function of the fraction of  values of $\delta$  for which a given significance could be achieved for different beam combinations and with improved neutrino energy resolution (Gaussian smearing). 
}}
\label{fig:frac}
\end{figure}

 The $\chi^2$ is computed as given in Eq.~\ref{chisq_mh} for a given set of true values by minimizing over the test parameters and this procedure is repeated for all possible true values 
  listed in Table~\ref{tab:parameters}. The shape of the sensitivity curves can be explained 
  using analytic form of probabilities~\cite{Masud:2016bvp,Masud:2016gcl}.
 \\

 \begin{figure}[hbt!]
\centering
\includegraphics[width= .8\textwidth] 
{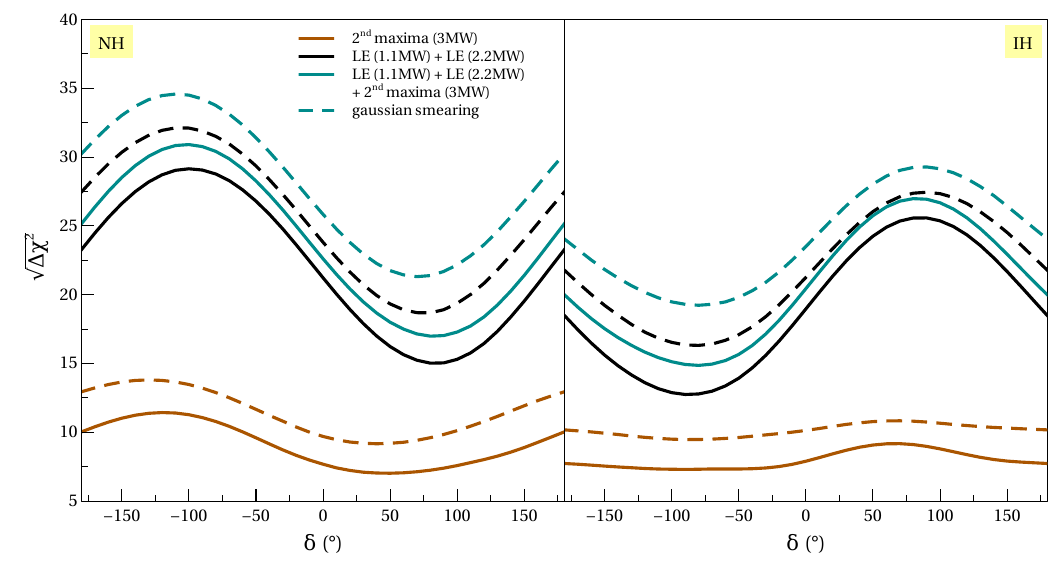}
    \caption{\footnotesize{Senstivity to the neutrino mass hierarchy   as a function of $\delta$ for various beam combinations and with improved neutrino energy resolution (Gaussian smearing).      
    }}
    \label{fig:masshir}
\end{figure}

    {\bf{Sensitivity to the octant of $\theta_{23}$ :-}}    
    It is  important   to determine the value of  $\sin^2  \tc$ with sufficient precision to determine the octant of $\tc$.
    A combination of $\nu_e$ appearance (which is sensitive to $\sin ^2 \tc$) and $\nu_\mu$ disappearance (sensitive to $\sin^2 2 \tc$) measurements would allow us to probe both maximal mixing and the octant of $\tc$. 
The $\Delta \chi^2$ is defined as~\cite{Acciarri:2015uup} 
\begin{equation}
\label{chisq_octant1}
\Delta \chi^2_{octant}  =
| \chi^2_{\tc^{test} > 45^\circ} - \chi^2_{\tc^{test} < 45^\circ} | 
\end{equation}
where the value of $\tc$ in the wrong octant is constrained only to have a value within the wrong octant (i.e., it is not required to have the same value of $\sin^2 2\tc$ as the true value).

\section{Results}
\label{results}

One of the primary objectives of DUNE is to ascertain whether CP is violated in the leptonic sector 
  within the context of the standard three flavour mixing scenario. In Fig.~\ref{fig:9m}, we depict the CP violation sensitivity  as a function of $\delta$ for different beam tune combinations. 
   The largest sensitivity to CP violation occurs at  $\delta = \pm \pi/2$. This is expected because the difference between the event rates at $\delta=\pm\pi/2$ (maximal CP violating values) and 
    $\delta_{}=0$ or $\pi$ (CP conserving values) is the largest. The  contribution from the second  
     maxima beam (brown curve) alone stays below $3\sigma$ level for all values of the CP phase even with the improved energy resolution. 
      The contribution from LE 1.1 MW and LE 2.2 MW beam taken 
     together is shown as a black curve and for the considered exposure (5+5 yrs), the sensitivity lies above $3\sigma$ level for {\hlpm{$\sim 74 \% ~(78\%)$}} of the possible values of $\delta$ for NH (IH).  
    The combination LE 1.1 MW and LE 2.2 MW beam along with $2^{nd}$ maxima beam (running concurrently with LE 2.2 MW) produces a
     {\sl{modest improvement in the sensitivity to CP violation}} near the maximal CP violating values. This holds irrespective of the choice of hierarchy. {{We depict the same information in Fig.~\ref{fig:frac} but as a function of fraction of values of $\delta$ that allow for discovery of CP violation at the corresponding significance. We note that improved energy reconstruction capabilities with better neutrino energy resolution could lead to much better significance (as shown by the dashed lines) especially in the neighbourhood of maximal CP violating values. 
     }}

\begin{figure}[htb!]
\centering
\includegraphics[width= .8\textwidth]{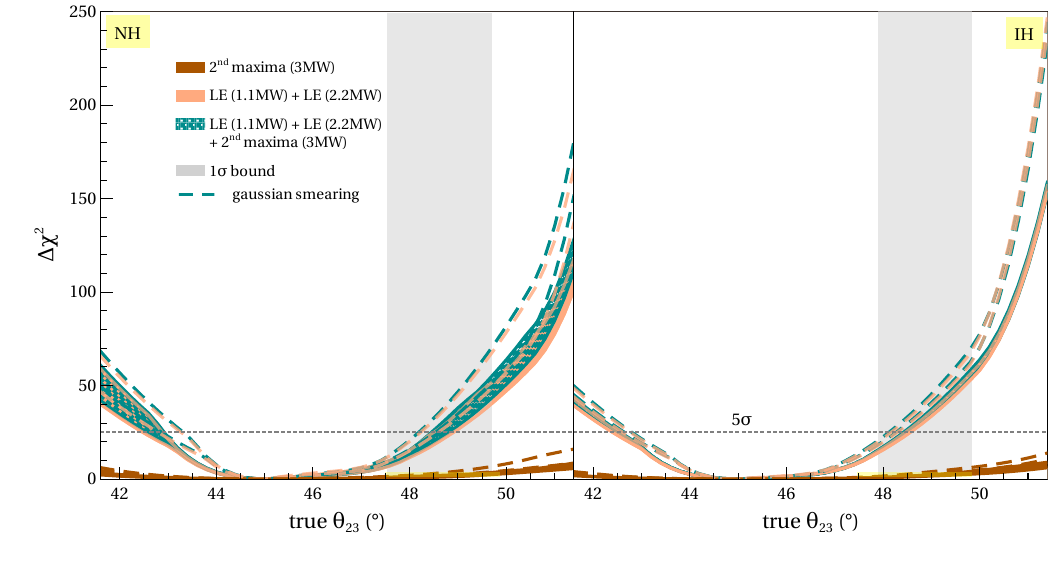}
    \caption{\footnotesize{{
    The significance with which DUNE can resolve the $\tc$ octant as a function of the true value of $\tc$ for different combinations of beams and with improved neutrino energy resolution (Gaussian smearing). 
    The shaded band around the curve represents the range in sensitivity due to potential variations  in the true value of $\delta$. 
    The grey shaded regions indicate the current $1\sigma$  bounds on the value of $\tc$ from a global fit.   
     }}}
    \label{octsen}
\end{figure}

In Fig.~\ref{fig:masshir}, we show   sensitivity to the MH  as a function of $\delta$. Here again, we get significantly better results for the LE beam running only as compared to the $2^{nd}$ maxima beam, when taken in isolation. However, when we consider the combination of beam tunes (LE 1.1 MW, 
LE 2.2 MW and $2^{nd}$ maxima), we notice a {\sl{modest overall improvement in the sensitivity to MH}} for all values of $\delta$. With the exposure considered, one can discern the MH definitively\,\footnote{{See~\cite{Qian:2012zn,Ciuffoli:2017ayi} for a discussion on statistical methods in estimation of the neutrino MH.}} 
for the LE beam combination as well as for the combination of LE with the $2^{nd}$ maxima. {{With better energy reconstruction capabilities, we find that the MH can be deciphered even better for all values of $\delta$.}}

\begin{figure}[htb!]
\centering
\includegraphics[width= .8\textwidth] 
{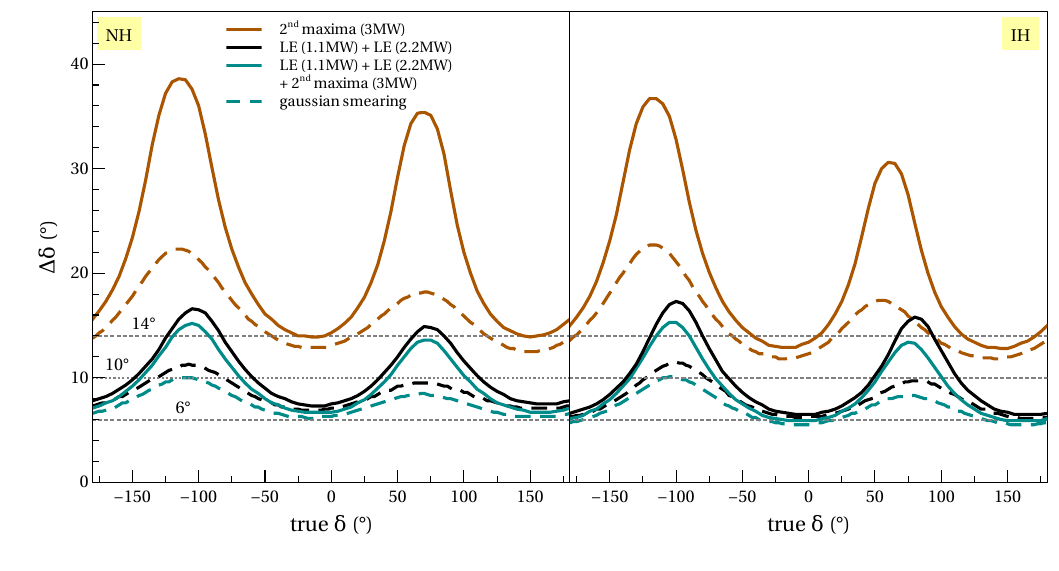}
    \caption{\footnotesize{Resolution on the measurement of $\delta$ as a function 
     of the true value of $\delta$ for various beam combinations and with improved neutrino energy resolution (Gaussian smearing).
 }}
    \label{cpres}
\end{figure} 
%
\begin{figure}[ht!]
\centering
\includegraphics[width=.8\textwidth]{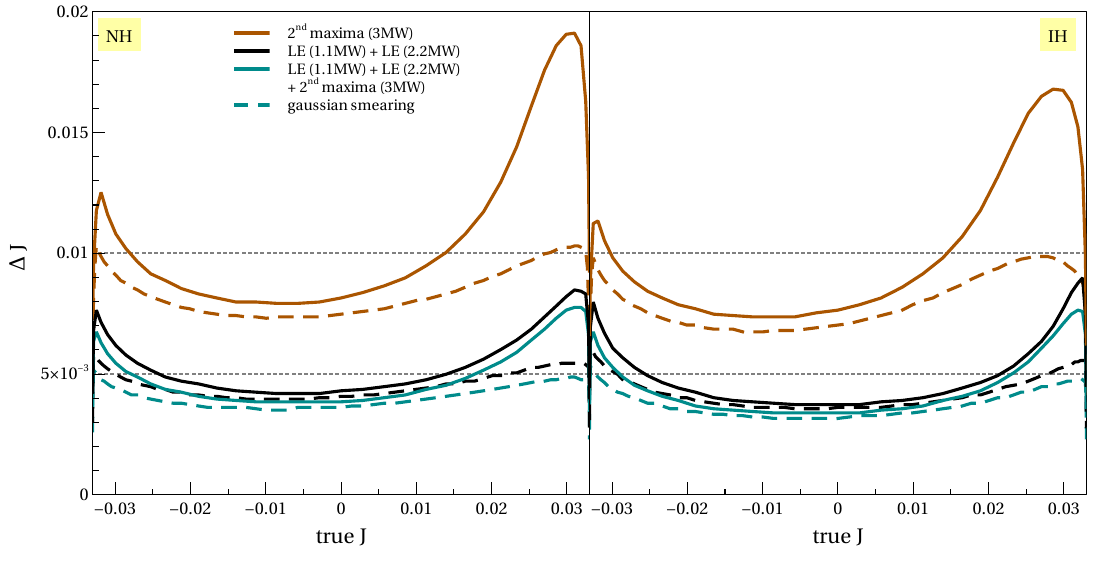}
    \caption{\footnotesize{{{{Resolution on the measurement of the Jarlskog factor, $J$ as a function 
     of the true value of $J$ for various beam combinations and with improved neutrino energy resolution (Gaussian smearing). }}}
 }}
    \label{jres}
\end{figure} 

DUNE will not only address questions pertaining to CP violation and neutrino MH, but also improve the precision on key parameters ($\sin^2 \theta_{23}$ and the octant of $\theta_{23}$, $\delta$, $\sin^2 2 \theta_{13}$ and $\Delta m^2_{31}$) entering the oscillation framework.  It is crucial to determine the value of $\sin^2 \tc$ with sufficient precision to determine the octant. The sensitivity of determining the octant  of $\tc$ as a function of true value of $\tc$ for different beam combinations is shown in Fig.~\ref{octsen} for NH and IH.   
{ {Adding the 8 GeV beam data to 80 GeV data at DUNE  {\sl{improves the sensitivity to resolution of the $\tc$ octant degeneracy}}.  
The width of each curve is due to the unknown CP phase and covers all possible true $\delta$ values. The grey shaded regions indicate the current $1\sigma$  bounds on the value of $\tc$ from a global fit~\cite{deSalas:2020pgw}. 
 In this fit, muon neutrino disappearance contributes to the precision on $\sin^2 2 \tc$ while the electron neutrino appearance data provides information on the $\tc$ octant.   A $5\sigma$ determination of the octant of $\tc$ will be possible for at least 90\% of true values of $\delta$ for $42.5^{\circ} < \tc < 49 ^\circ$ for NH. Most of the sensitivity comes from the LE 
  data because the $2^{nd}$ maximum does not provide any added sensitivity to the octant determination. 
  {{We show the impact of improved energy reconstruction capabilities on the octant of $\tc$ as dashed lines.}} 

Fig.~\ref{cpres} shows the one-dimensional resolution on the measurement of $\delta$ (see~\cite{Coloma:2012wq,Coloma:2014kca} for details) the 3 beam combinations considered: LE beam only 5 yrs at 1.1 MW + 5 yrs at 2.2 MW; the 3MW $2^{nd}$ maxima beam only, and the 10 years of DUNE running assuming the 3MW $2^{nd}$ maxima beam runs concurrently with the 2.2 MW LE beam for 5 of the 10 years. Running with the 3MW $2^{nd}$ maxima beam in conjunction with standard LE beam running at DUNE 
leads to {\sl{some improvement in $\delta$ resolution}} for almost all true values of $\delta$, irrespective of the choice of hierarchy.  
{{Improved energy reconstruction capabilities lead to significantly better $\delta$ resolution especially near the maximal CP violating values. It is worth pointing out that with the considered beam combination and the improved neutrino energy resolution, $\delta$ can be resolved within $6-10^\circ$ for all values of $\delta$.}}  {\hlpm{The resolution of the CP phase at $\delta=\pi/2$ lies between $\sim 11.8 - 14.9^\circ$  ($\sim 8 - 9.6^\circ$)  for nominal (improved) detector resolution. It can be noted that the $2^{nd}$ maxima beam has a bigger impact on $\delta$ resolution for the case of IH. 
 At $\delta=0$, we obtain ${\Delta \delta} \sim  5.9 - 7.5 ^\circ$  $(\sim 5.5 - 7.1^\circ)$ with nominal (improved) detector resolution.     }}

\begin{figure}[htb!]
\centering
\includegraphics[width= .8\textwidth] 
{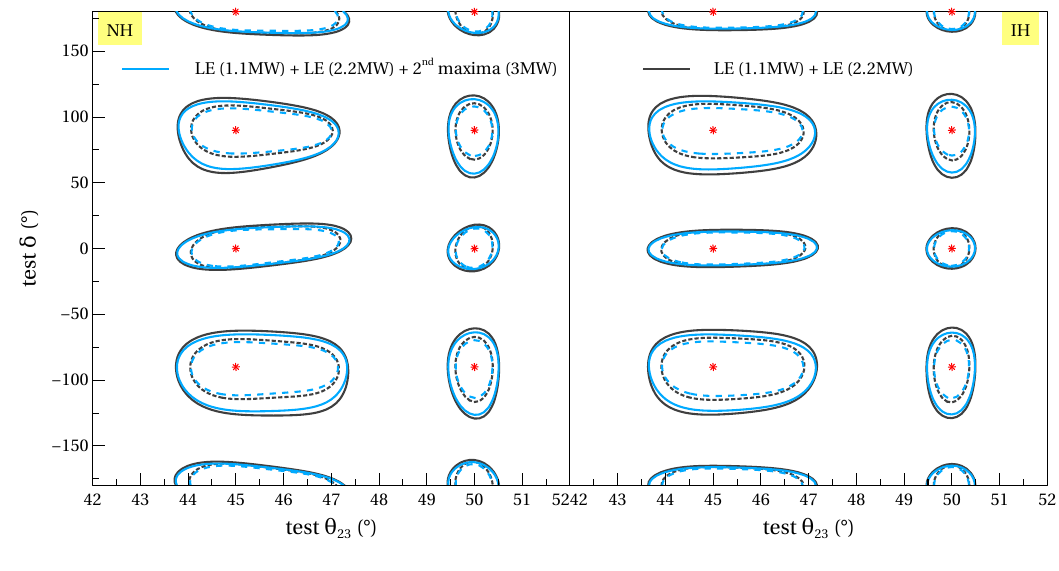}
    \caption{\footnotesize{\footnotesize{$1\sigma$ contour plots  from a two-dimensional fit for $\tc$ and $\delta$ for the different beam tune combinations with DUNE configuration files (solid lines) and improved neutrino energy resolution (dashed lines).  
     Fit results for various possible true values of  $\delta = 0, \pm
     \pi/2, \pi$ are shown.}  
     }}
    \label{cont1signh}
\end{figure}

{{It is important to note that our inferences pertaining to the CP violating phase, $\delta$ in neutrino oscillations depend on the specific parameterization of the mixing matrix (Eq.~\ref{u})
and different parameterizations   could result in a completely different value of $\delta$~\cite{Denton:2020igp}.  However, the Jarlskog factor is independent of any parameterization or phase convention, and hence an invariant. 
The  resolution on the measurement of the Jarlskog invariant, $J$  is  shown in Fig.~\ref{jres}. %
Additional running with the $2^{nd}$ maxima beam using the default neutrino energy smearing produces very modest improvements to the resolution of $J$ over all values. By contrast, improvements in the neutrino energy resolution produce significant improvements in $\Delta J$ at values of $J$ near 0.033. The combination of running in the $2^{nd}$ maxima beam and improved energy resolution for 10 years of DUNE operation (assuming 8GeV @ 3MW runs concurrently with 80 GeV @ 2.2MW  with the PIP-III SRF linac option) could yield a resolution of $5 \times 10^{-3}$ or better for all values of $J$.
}}}}

The $1\sigma$ contours from a two-dimensional fit to $\tc$ and $\delta$ with external constraint 
on $\theta_{13}$ for the combinations considered in the present work are shown in Fig.~\ref{cont1signh}.  {{Improved energy reconstruction capabilities lead to shrinking of these contours.}}
 Additionally,  data from the LE and $2^{nd}$ maxima beam place independent constraints on other 
  neutrino oscillation parameters as well.

\section{Conclusion}
\label{conclude}

The present study highlights the theoretical importance of the second oscillation maximum in the $\nu_\mu \to \nu_e$ oscillation channel and its impact on addressing the current unknowns in neutrino oscillations. The international accelerator-based long baseline 
neutrino program aims to constrain the parameters of the mixing matrix (or find deviations from the standard paradigm) through extremely precise measurements  of flavour oscillations with systematic uncertanities well within the percent level. The idea of exploring physics at the second (and higher) maxima of $P_{\mu e}$ in the context of various   experiments 
 is of paramount importance as it allows for a complete underpinning of the standard three-flavour oscillation framework - particularly the unknown CP violating phase - since it allows us to study a wide $L/E$ range with fixed baseline experiments.
 
  In recent times, there has been an upsurge in activity relating to analyzing the role of second oscillation maximum in the context of ESS$\nu$SB (which  covers the second oscillation maximum using a 2 GeV $5$ MW proton linac as a neutrino source and a Water Cherenkov detector placed at $\sim 500$ km away) and T2HKK both within and beyond the standard three flavour mixing 
  paradigm~\cite{Baussan:2013zcy,Wildner:2015yaa,Blennow:2019bvl,Agarwalla_2014,Chakraborty_2019,Ghosh:2019sfi,Ghosh:2020vly,Choubey:2020dhw,Chakraborty:2020cfu}. The prospect of precision measurement of $\delta$ at MOMENT ($L = 150$ km) has also been studied in~\cite{Cao_2014,Tang:2019wsv}.
In addition, the T2HKK~\cite{Hagiwara_2008} proposal measures the first and the second oscillation maxima  with two detectors located at different sites.   Therefore, it is  timely to   carry out a comprehensive study investigating the role of second oscillation maximum in the context of DUNE. 

The usefulness of observing the second oscillation maximum can be understood in terms of an argument based on CP asymmetries. In order to observe a signal for CP violation in the leptonic sector, one needs to measure  the
 CP conjugate channels ($\nu_\mu \to \nu_e$ and $\bar \nu_\mu \to \bar\nu_e$).  The interference term 
  has the CP violating parameter $\delta$ and should be larger (ideally) compared to the other two terms (solar and atmospheric). For the measured value of $\theta_{13}$, the interference term is 
    large compared to the other two terms.  The CP   asymmetry is larger at the second oscillation maximum since $A^{CP}_{\mu e} \simeq 0.75 \sin \delta $ (at $L/E \simeq 1500$ km/GeV) and $A^{CP}_{\mu e} \simeq 0.3 \sin \delta $ 
    (at $L/E \simeq 500$ km/GeV). This implies  significantly higher sensitivity to observe CP violation. Now, if  neutrino energy      is held fixed, the baseline has to be about three times larger than that required for the first oscillation maximum. But, this leads to a reduction in statistics by an order of magnitude. However, the other potentially viable option is to use 
     a very intense neutrino beam from a multi-MW proton beam.  Our experimental setup comprises of LBNF/DUNE using a neutrino beam covering the $2^{nd}$ maxima  in conjunction with the standard CP optimized wide-band beam currently under design~\cite{Acciarri:2015uup}. The neutrino beam at the second maxima is generated using a 3 MW, 8 GeV proton beam which could be realized by the PIP-III SRF linac option~\cite{pip}. We note that the default CP optimized wide-band beam for DUNE already covers a significant portion of the $2^{nd}$ oscillation maximum and offers the capability of precision measurement using the shape of the observed oscillation over a wide range of $L/E$. The addition of the $2^{nd}$ maxima beam running enhances the flux of neutrinos at the $2^{nd}$ maxima without the penalty of the background feed-down from higher energy neutrinos (in particular the backgrounds from NC and $\nu_\tau$).

{\hlpm{In Table~\ref{tab:final}, we   list the expected improvement in sensitivity to the different unknowns with the   beam and runtime combinations (a) and (c)   mentioned in Sec.~\ref{event}.}}
%
\begin{table}[hbt!]
{
\scriptsize
{
\centering
\begin{tabular}{|l | c | c | c |  c | }
\hline
 {\bf{Sensitivity to}} & 
 \multicolumn{2}{c|}{\bf{(a) LE, 1.1 MW + LE, 2.2 MW}}    & 
 \multicolumn{2}{c|}{\bf{(c) LE, 1.1 MW + LE, 2.2 MW}} \\
{\bf{}} & 
 \multicolumn{2}{c|}{\bf{}}    & 
 \multicolumn{2}{c|}{\bf{+ $2^{nd}$
maxima, 3 MW}}  \\  
  {\bf{}} & 
 \multicolumn{2}{c|}{\bf{}}    & 
 \multicolumn{2}{c|}{\bf{}}  \\  
\cline{2-5} 
&
 {{NH}} & {{IH}} & {{NH}} & {{IH}}   \\
\hline
{\sl{(i) CP violation}} &&&&  \\
$[\sigma]$ at $\delta = \pi/2$              & 6.9 (8.9) & 8.5  (9.9) & 7.6  (9.9) & 9.0 (10.8) \\
\hline
{\sl{(ii) CP fraction}} &&&&\\
$f(\ge 3 \sigma)$            & 0.74 (0.75) & 0.78 (0.79) & 0.77 (0.78) & 0.80 (0.81)  \\
$f(\ge 5 \sigma)$            & 0.54 (0.58) & 0.61 (0.64) & 0.59 (0.63) & 0.65 (0.68)  \\
 \hline
{\sl{(iii) MH}} &&& &\\
$[\sqrt{\Delta \chi^2}]$ at $\delta= \pi/2$ & 15.0 (18.9) & 25.6 (27.5) & 17.0 (21.7) & 26.9 (29.3)  \\
 \hline
{\sl{(iv) Octant of $\theta_{23}$}} &&&&\\
$[{\Delta \chi^2}]$  at $\theta_{23}= 48.8^\circ$ & 30.8 (39.3) & 31.8 (39.0) & 33.2 (40.8) & 32.8 (40.3)  \\
\hline
{\sl{(v) $\delta$ resolution}} &&&&\\
$[^\circ]$ at $\delta= 0$     & 7.5 (7.1) & 6.5 (6.3) & 6.7 (6.3) & 5.9 (5.5)  \\
$[^\circ]$ at $\delta= \pi/2$ & 13.3 (9) & {\bf{14.9}} (9.6) & 11.8 (8) & {\bf{11.9}} (8) \\
\hline
{\sl{(vi) $J$ resolution}} &&&&\\
$[\times 10^{-3}]$    at   $J =  0 $                & 4.3 (4.1)  & 3.7 (3.6)  & 3.9 (3.6) & 3.4 (3.2)  \\
$[\times 10^{-3}]$    at  $J =  0.033$ & 4.0 (2.7)  & 4.5 (2.8)  & 3.6 (2.4)  & 3.6 (2.4)  \\
\hline
\end{tabular}
\caption{\label{tab:final} \footnotesize{{\hlpm{Expected sensitivity to the different unknowns with the considered beam combinations (a) and (c) 
  mentioned in Sec.~\ref{event}.  We consider a runtime of 5 yr in LE, 1.1 MW and 5 yr in LE, 2.2 MW
  distributed equally in  neutrino and antineutrino modes for beam option (a). 
 We consider a runtime of 5 yr in LE, 1.1 MW,  
   5 yr in LE, 2.2 MW and 5 yr with $2^{nd}$ maxima, 3 MW 
   distributed equally in neutrino and antineutrino modes for beam option (c). 
  The numbers in brackets correspond to  scenario with improved energy reconstruction capabilities for the considered beam combinations. }}
}}}}
\end{table}
We summarize the main  results of our sensitivity studies contained in Figs.~\ref{fig:9m}-\ref{cont1signh} as follows. 

\begin{itemize}

\item Using the default DUNE configuration files~\cite{Alion:2016uaj} and adding the $2^{nd}$ maxima beam running concurrently with LE 2.2 MW beam (PIP-III SRF linac option), we find {\sl{modest}} improvement in sensitivity to CP violation, MH and the octant of $\tc$ (see Figs.~\ref{fig:9m},  \ref{fig:frac}, \ref{fig:masshir} and \ref{octsen}).  {\hlpm{As can be seen from Fig.~\ref{cpres}, the considered beam combination leads to a {\sl{modest improvement  in $\delta$ resolution of $\sim 1-2^{\circ}$ in the vicinity of maximal CP violating values ($\delta = \pm \pi/2$)}} for the case of NH. For IH, we get {\sl{slightly larger improvement in $\delta$ resolution $\sim 3^\circ$}} at $\delta = \pi/2$.  }}

 \item {\hlpm{Improved energy reconstruction capabilities lead to {\sl{significantly better}} sensitivities to CP violation, MH and octant of $\tc$ (see Figs.~\ref{fig:9m}, \ref{fig:frac}, \ref{fig:masshir} and \ref{octsen}) for values of $\delta$ near maximal. The improved resolution of $\delta$ from improved energy reconstruction significantly outperforms the gains from running in the $2^{nd}$ maxima beam in the vicinity of maximal CP violating values (see Fig.~\ref{cpres}). }}

\item 
{\hlpm{Fig.~\ref{fig:9m} depicts the  sensitivity to CP violation as a function of  $\delta$.    The CP violation sensitivity at $\delta=\pi/2$ approaches $\sim 6.9 - 9 \sigma$  ($\sim 8.9 -10.8 \sigma$)  for nominal (improved) detector resolution.}}

 \item  {\hlpm{Fig.~\ref{fig:frac} shows the  sensitivity to CP violation  as a function of fraction of  values of $\delta$  for which a given significance could be achieved.   
The $3\sigma$ ($5\sigma$)  discovery of CP violation can be achieved for {{$\sim 74-80\%$ ($\sim 54-65\%$)}} values of the CP phase for nominal  detector resolution.}}
  {\hlpm{With improved energy reconstruction capabilities, $3\sigma$ ($5\sigma$)  discovery of CP violation can be achieved for $\sim 75-81\%$ ($\sim 58-68\%$) values of the CP phase.}}  

\item {\hlpm{Fig.~\ref{fig:masshir} depicts the  sensitivity to MH   as a function of  $\delta$.  We note that for $\delta=\pi/2$, the $\sqrt{\Delta \chi^2} $ corresponding to the MH sensitivity  lies within  $ \sim 15 - 26.9$  ($\sim 18.9 - 29.3$)  for nominal (improved) detector resolution.}}

\item {\hlpm{The sensitivity of determining the octant  of $\tc$ as a function of true value of $\tc$ for different beam combinations is shown in Fig.~\ref{octsen} for NH and IH.   A $5\sigma$ determination of the octant of $\tc$ will be possible for at least 90\% of true values of $\delta$ for $42.5^{\circ} < \tc < 49 ^\circ$ for NH.  We note that for $\theta_{23}=48.8^\circ$, the ${\Delta \chi^2} $ corresponding to the octant sensitivity  lies within  $ \sim 30.8 - 33.2$  ($\sim 39 - 40.8$)  for nominal (improved) detector resolution.}}

 \item  {\hlpm{One of our key results pertains to the the improvement in $\delta$ resolution. It is shown in Fig.~\ref{cpres} that $\delta$ can be {\sl{resolved better than $\sim 10^\circ$}} for all values of $\delta$ with   improved energy resolution and additional running with the  $2^{nd}$ maxima beam. {\hlpm{The resolution of the CP phase at $\delta=\pi/2$ lies between $\sim 11.8 - 14.9^\circ$  ($\sim 8 - 9.6^\circ$)  for nominal (improved) detector resolution. It can be noted that the $2^{nd}$ maxima beam has a bigger impact on $\delta$ resolution for the case of IH. 
 At $\delta=0$, we obtain ${\Delta \delta} \sim  5.9 - 7.5 ^\circ$  $(\sim 5.5 - 7.1^\circ)$ with nominal (improved) detector resolution.     }}}}  
 
 \item   We also deduce implications on the measurement of the Jarlskog invariant $J$ which is a parameterization independent quantity.  The improvement on the resolution on the Jarlskog invariant $J$ as a function of $J$ due to improved energy resolution and additional running in the $2^{nd}$ maxima beam is depicted in Fig.~\ref{jres}. Additional running with the $2^{nd}$ maxima beam using the default neutrino energy smearing produces very modest improvements to the resolution of $J$ over all values. By contrast, improvements in the neutrino energy resolution produce significant improvements in $\Delta J$ at values of $J$ near 0.033 {\hlpm{(corresponding to $\delta = \pi/2$)}}.  {\hlpm{${\Delta J} \simeq 3.6 - 4.5 \times 10^{-3} $  ($\sim 2.4 - 2.8 \times 10^{-3} $)  for nominal (improved) detector resolution at $\delta=\pi/2$ irrespective of the  hierarchy. 
 ${\Delta J} \simeq 3.4 - 4.3 \times 10^{-3} $  ($\sim 3.2 - 4.1 \times 10^{-3} $)  for nominal (improved) detector resolution at $\delta=0$. 
 }} The combination of running in the $2^{nd}$ maxima beam and improved energy resolution for 10 years of DUNE operation (assuming 8GeV @ 3MW runs concurrently with 80 GeV @ 2.2MW  with the PIP-III SRF linac option) could yield a resolution of $5 \times 10^{-3}$ or better for all values of $J$.
 
 \item Finally, we show the $1\sigma$ contours from a two-dimensional fit to $\tc$ and $\delta$ for the beamtune combinations considered in the present work (see Fig.~\ref{cont1signh}) both  with DUNE configuration files~\cite{Alion:2016uaj} and with improved energy reconstruction capabilities. 
 
\end{itemize}

We conclude that the default wide-band CP optimized beam for DUNE already offers an excellent opportunity to probe physics in the vicinity of the $2^{nd}$ oscillation maxima. Improved neutrino energy resolution - coupled with very low uncertainties on the energy reconstruction - is needed to fully realize this opportunity. The improved neutrino energy resolution could be achieved with further improvements in LArTPC reconstruction based on the latest DUNE studies and MicroBooNE data. The PIP-III upgrade with a pulsed SRF linac offers the opportunity for a multi-MW 8 GeV beam to DUNE with flux covering only the region of the $2^{nd}$ oscillation maxima, thus providing an opportunity for increasing the flux at the  $2^{nd}$ maxima without the additional background from feed-down from higher energy neutrinos. Such a beam, running concurrently with the 2.2 MW 80 GeV beam from PIP-III can further improve sensitivity to the 3-flavor oscillation parameters, but we find the gain is modest.

\section*{Acknowledgements} 
{{It is a pleasure to thank Peter Denton for  insightful discussions and Enrique Fernandez-Martinez for  email communication in connection with Ref~\cite{DeRomeri:2016qwo} after the preprint  appeared on the arXiv. }}
  We  acknowledge the use of HPC cluster at SPS, JNU funded by DST-FIST for numerical computations.    
JR and SS thank the University Grants Commission for financial support in the form of research fellowships.
This material is based upon work supported by the U.S. Department of Energy, Office of Science, Office of High Energy
Physics under contract number DE-SC0012704; the Indian funding from
University Grants Commission under the second phase of University with
Potential of Excellence (UPE II) at JNU and  Department of Science and Technology under  
 DST-PURSE at JNU.  This work has received partial funding from the European Union's Horizon 2020 research and innovation programme under the Marie Skodowska-Curie grant agreement No 690575 and 674896. {{PM would like to thank the particle physics group at Brookhaven National Laboratory for the warm hospitality during the early stages of this work.}}

\bibliographystyle{apsrev}
\bibliography{references_2020}

\end{document}